\documentclass[manuscript,screen]{acmart}
\usepackage{xcolor} %NEW
\usepackage[symbol]{footmisc} % NEW
\usepackage{listings}
\usepackage{multirow}
\usepackage[normalem]{ulem}
\usepackage{pdflscape}
\usepackage{tablefootnote}

\usepackage{lscape}

\definecolor{codegreen}{rgb}{0,0.6,0}
\definecolor{codegray}{rgb}{0.5,0.5,0.5}
\definecolor{codepurple}{rgb}{0.58,0,0.82}
\definecolor{backcolour}{rgb}{0.95,0.95,0.92}

\lstdefinestyle{pythonstyle}{
	language        = Python,
    backgroundcolor=\color{backcolour},   
    commentstyle=\color{codegreen},
    keywordstyle=\color{magenta},
    numberstyle=\tiny\color{codegray},
    stringstyle=\color{codepurple},
    basicstyle=\scriptsize\ttfamily\linespread{0.8},
    breakatwhitespace=false,         
    breaklines=true,                 
    keepspaces=true,                 
    numbers=left,                    
    numbersep=5pt,                  
    showspaces=false,                
    showstringspaces=false,
    showtabs=false,                  
    tabsize=2
}
\AtBeginDocument{%
  }

\setcopyright{acmcopyright}
\copyrightyear{2023}
\acmYear{2023}

\acmDOI{XXXXXXX.XXXXXXX}
\acmJournal{CSUR}

\begin{document}

\title[Landscape of High-performance Python]{Landscape of High-performance Python to Develop Data Science and Machine Learning Applications}

\author{Oscar Castro}
\orcid{https://orcid.org/0000-0003-4025-7903}
\authornote{Oscar Castro is now with University of Luxembourg}
\email{oscar.castro@list.lu}

\author{Pierrick Bruneau}
\orcid{https://orcid.org/0000-0002-7725-512X}
\email{pierrick.bruneau@list.lu}

\author{Jean-S\'ebastien Sottet}
\orcid{https://orcid.org/0000-0002-3071-6371}
\email{jean-sebastien.sottet@list.lu}
\affiliation{%
  \institution{Luxembourg Institute of Science and Technology}
  \streetaddress{5 avenue des Hauts Fourneaux}
  \city{Esch/Alzette}
  \country{Luxembourg}
}
\author{Dario Torregrossa}
\orcid{https://orcid.org/0000-0002-5863-1628}
\email{dario_torregrossa@goodyear.com}
\affiliation{%
  \institution{Goodyear Innovation Center}
  \streetaddress{Avenue Gordon Smith}
  \city{Colmar-Berg}
  \country{Luxembourg}}

\renewcommand{\shortauthors}{Castro et al.}

\begin{abstract}
	Python has become the prime language for application development in the Data Science and Machine Learning domains. However, data scientists are not necessarily experienced programmers. While Python lets them quickly implement their algorithms, when moving at scale, computation efficiency becomes inevitable. Thus, harnessing high-performance devices such as multicore processors and Graphical Processing Units (GPUs) to their potential is generally not trivial. The present narrative survey was thought as a reference document for such practitioners to help them make their way in the wealth of tools and techniques available for the Python language. Our document revolves around user scenarios, which are meant to cover most situations they may face. We believe that this document may also be of practical use to tool developers, who may use our work to identify potential lacks in existing tools and help them motivate their contributions. 
\end{abstract}




\maketitle

\section{Introduction}

Python is one of the most used computer programming language nowadays:  it is ranked in the first position on the PYPL (PopularitY of Programming Language) index \cite{PYPL2021} and first position on the TIOBE index \cite{TIOBE2021} in 2022. 

It is intensively used in the growing domains of Data Science (DS), scientific computation, data analytics, and Machine Learning (ML). It is used as the successor of the many data-centric and scientific computation programming languages such as R, Fortran, and Matlab. One of the main reasons behind this success in data science stands on its many DS and ML focused libraries such as NumPy, Pandas, TensorFlow, Scikit-learn, SciPy, and MatplotLib. Given the amount of data being collected and processed within the DS and ML contexts, most Python high-performance libraries have been developed outside Python by using statically typed languages such as C++, Fortran, and/or CUDA.

The main reasons that explain the fact that libraries are developed outside Python are the slow performances of the Python interpreter. Ismail and Suh \cite{ismail_quantitative_2018} studied in detail the overheads coming with Python code execution. First, as is generally true of interpreted languages, it is slower than running compiled code. Indeed, like most interpreted languages (e.g., Java), Python programs are translated to bytecode before execution by a virtual machine. An additional inherent inefficiency comes with Python due to its dynamic object typing system. Importantly, Ismail and Suh identify that C function calls from the interpreter, invoked when calling a compiled library function, yield significant overheads. A disposable execution environment then must be set up and cleaned, which brings a constant per-instruction overhead.

The default implementation, CPython, uses the Global Interpreter Lock (GIL). The GIL offers some safety mechanisms for concurrent accesses, in return it prevents multi-threading: the interpreter executes only a single thread within a single CPython process. The aim of the GIL is to simplify the implementation by making the object model safe against concurrent access. This means that CPython executes CPU-bound code in a single thread. In addition, evaluations showed that CPython exhibits poor instruction-level parallelism in this context \cite{ismail_quantitative_2018}. As a result, in terms of efficiency, Python is not doing well if compared with other languages. Therefore, there are different tools available to improve the performance of programs built in Python.

The objective of this review article is to provide an organized landscape of Python high-performance tools. We can relate to surveys considering the acceleration and optimization of communications in the context of high-performance computing \cite{rico-gallego_survey_2019}, or improving compilation to accelerate code execution \cite{bacon_compiler_1994}. Parallelization of inputs and outputs in the context of high-performance computing \cite{boito_checkpoint_2018}, software optimization in view of energy efficiency of embedded systems \cite{paulino_improving_2020} and GPU acceleration in the context of a specific Machine Learning task, for example Frequent Itemsets Mining \cite{bustio-martinez_fpgagpu-based_2021} are also closely related domains. But to our knowledge, we propose the first survey of high-performance and code acceleration tools restricted to the Python programming language. An originality of our approach is also to organize our work according to user scenarios. As such, it aims at identifying what are the different categories of approaches used for Python code acceleration regarding different prototypical DS and ML practitioners profiles. This will facilitate adoption by practitioners looking for guidance, while providing principal axes for disclosing our results. 
%We begin with a short introduction to Python and its evolution, its different back-ends (interpreter/compiler).
We begin by introducing our method and approach regarding this survey, notably our specific viewpoint based on practitioner profiles and scenarios. Specifically, we motivate the definition of three user profiles, designed to cover most situations and needs faced in the practice of DS and ML. We then disclose the details of the associated search strategy. In three sections, each mirroring the respective user profile, we group Python tools, primarily according to their relevance to the identified profiles, then by the concepts and techniques involved in view to improve Python performance. As this may not lead to a strict taxonomy, a versatile tool which may be applicable in more than one profile will be described in the section most closely matching its main usage scenario, as reflected in \emph{quick-start} sections and tutorials commonly seen for the tool. References will be used from other sections as needed. Within and between sections, besides describing the tools and providing quantitative and qualitative information about them, care is taken to emphasize similarities, complementarities, and potential conflicts between tools and approaches. We close the document with a discussion section which summarizes our findings and delineates some important lessons learned.  

\section{Method}
\label{sec:method}
This article presents a narrative review covering the domain of the acceleration of Python program execution. Thus, it aims at providing practitioners with an overview of what is the current state of the art in high-performance Python programming, notably through parallelization, distributed execution, and code transformation.

This review is initially motivated by our practice of Python programming in the domains of DS and ML, where improving performance is critical to obtain timely results \cite{castro_parallelization_2022}.
This initial work led us to realize that a systematic review focused on high-performance Python with the practice of DS and ML in mind is lacking from the current literature. As our work is driven by pragmatic concerns, we believe that a narrative review is the most suitable format.

\subsection{Approach and context}
As with any narrative review, we will do a qualitative evaluation of the diverse extant approaches in the domain of performance improvement of Python programs.
To lead this survey, the authors relied on their expertise, in the related fields, including industrial experience and research communities: ML, DS, software engineering. 
Their different backgrounds and institutions, as well as years of experience in the aforementioned fields, helped considering both academical and practical point of view, in a way as broad and relevant as possible.

However, the landscape of tools and techniques in this scope is very diverse, and may be distinguished according to a large number of facets (e.g., level of automation, close ties with a peculiar DS task, expected amount of effort to put in use).
As a result, there is no obvious hierarchy in these facets which would drive the structure of the taxonomy presented in this paper. Instead, we focus on three common usage scenarios, which are meant to be mostly mutually exclusive, while covering most situations met in practice by data scientists.

\begin{itemize}
	
	\item The developed algorithm may involve some non-standard data structure, such as a special kind of knowledge graph. To avoid the effort of searching for an existing library that can be adapted to fit the requirements, a pure prototypical Python algorithm may have been developed by a data scientist to solve a theoretical or practical problem. However, upon validation, the algorithm's performance significantly degrades when applied to larger datasets. As a result, there is a need to explore alternative approaches for more efficient computations.
	
	\item Most commonly, DS practitioners face situations close to canonical problem involving standard data structures such as numerical matrices or graphs. Practitioners will then design an algorithm to solve the  problem and implement it using popular numerical Python libraries such as Numpy or Pandas. After validating the algorithm, they need to apply it to larger data sets, but runtime becomes excessive.

	\item Finally, the algorithm may still be only on paper, and instead of boldly starting implementing it in vanilla Python, the data scientist may look for the right library or framework to directly maximize computational efficiency at implementation time, even if it involves learning to master a Domain Specific Language (DSL) or non-standard constructs.
\end{itemize}

In these three scenarios performance is sought, but from differing starting points, and with variable will to invest in mastering sophisticated tools. For instance, in the two first scenarios, implementations are already developed, and the goal is to look for cost-effective solutions to scale them up. In contrast, the third scenario is bound to a longer-term view, where the practitioner will tolerate to invest in the most appropriate tool from the start. 

Given this context, the objective of this survey is to answer the following question: \emph{which approaches can be used to improve Python execution performance in the context of one of these three scenarios?}

\subsection{Search Strategy}

Our search strategy started by using Google Scholar, which is a search engine that indexes metadata or even the whole text of scientific publications in multiple scientific repositories. In order to reflect our inclusion and exclusion criteria (see Table \ref{tab:searchplan}), we used the keyword \emph{Python} combined to \emph{Data Science} or \emph{Machine Learning}, as well as other keywords associated to high-performance computation and code acceleration (see also Table \ref{tab:searchplan}). The first seed of relevant papers thus collected could point us out to other important works by inspecting the cited references. We also looked at implementations and repositories referred to in this set of selected papers. We also covered social networks and forums commonly used by data science practitioners and software engineers for exchanging and sharing, such as Reddit, Stack Overflow, Kaggle and Data Science Central. We applied our search strategy on reference code repositories and Python package indexes as well, mostly GitHub and PyPI. Scientific publications were primarily filtered by date between 2018 and 2022. Nonetheless, we also review some tools dated before that time frame if they have very high relevance to our scope or if they have been actively maintained. The summary of the search strategy is presented in Table \ref{tab:searchplan}.

As we have stated beforehand, we focus on performance improvement in the context of three identified profiles associated to the above defined scenarios. For the second scenario, in the associated section we motivate the subset of libraries under our focus and describe how the search strategy was specifically amended in that case.

\begin{table}[h]\centering\scriptsize
	\begin{tabular}{lp{7.5cm}}
		
		\toprule
		
		\textbf{Source} & 
		\textbf{Criteria} \\ 

		\midrule    
		Database &  Web-based resources: Google Scholar, Github, PyPI, Reddit, Stack Overflow, Data Science Central\\ \hline
		Date of publication &  2018 - 2022 (for scientific publications) \\ \hline
		Keywords &  (\emph{high-performance} or \emph{parallelization} or \emph{multicore} or \emph{GPU} or \emph{performance improvement} or \emph{performance enhancement} or \emph{acceleration} or \emph{compilation} or \emph{transformation}) combined to \newline
		(\emph{Python} and (\emph{machine learning} or \emph{data science}))
		\\ \hline
		Language &  English \\ \hline		
		Type of publication &  Conference proceedings \newline
		Journal articles \newline
		PyPI documentation \newline
		GitHub vignettes \newline
		Social media posts \newline
		Web news and articles
		\\ \hline				
		Inclusion criteria &  Tools/Packages aimed at increasing the performance of Python code for ML and DS directly or indirectly. \\ \hline				
		Exclusion criteria &  Tools/packages outside Python. \newline
		Python packages focusing on one application use case. \newline
		Packages dedicated to a specific type of hardware other than CPU and GPU.
		\\ \hline								

		\bottomrule
	\end{tabular}
	\caption{Search plan summary}
	\label{tab:searchplan}
\end{table}

\subsection{Inclusion and Exclusion Criteria}

The selected publications and tools must propose some performance improvement of the execution of Python programs that directly impacts Data Science or Machine learning tasks. Some tools with a wider scope may be included, provided that they clearly enable to accelerate the execution of DS or ML tasks, or relevant function in tier DS or ML packages.

We restricted the resulting scope to the Python realm. We are aware that many high-performance libraries and research exist independently of Python, but our choice is motivated by the relative monopoly of Python in the domains of DS and ML. Some of the approaches, mainly developed in C/C++ but accessible thanks to wrappers in Python, are obviously considered.
While we focus on recent tools and contributions, we also report about seminal work if its lineage can be directly related to recent practice.
A high number of works are from 2015 and later due to the relative acceleration of contributions in ML and DS recently.
The diversity implied by our scenarios led us to consider all levels of granularity in terms of enhancing the performance of Python code: from very general code transformation approaches to numerical libraries widely used in the DS domain.

We focused on the default CPython interpreter, and thus did not consider alternative Python interpreters (e.g., Pyston \cite{pyston2022}) intentionally. We address this point in a specific paragraph in the discussion (Section \ref{sec:interpreters}).
Besides pure code optimization, we will also consider approaches exploiting multiple CPUs (and CPU cores), as well as GPUs, the latter being widely used in ML. Nevertheless, we did not dig into application specific or dedicated processors like Tensor Processing Units (TPUs) and Field-Programmable Gate Arrays (FPGAs).

\section{Pure Python performance improvement} 
\label{sec:generalpurpose}
In this section, we focus on tools and approaches that support acceleration of code in which the computationally intensive parts rely only on the default Python distribution (also called vanilla Python).
In the context of our scenarios, this can be because the modelling of the problem at hand is not standard and thus not necessarily compliant with existing numerical computation libraries (e.g., custom knowledge graph). This can be also because the practitioner is more comfortable with vanilla Python for working on an implementation which sticks to some algorithmic formalism in the literature.

Acceleration approaches in this section will generally target Python at large, beyond traditional DS use cases. In this scenario, we thus assume that a working version of the code addressing the problem of the practitioner has already been developed. It works on small samples of data, but now needs to be accelerated to be scaled up to larger amounts of data. 
DS and ML tasks often involve loops performing the same operation on many chunks of data (e.g., when independently processing images or database records). The seminal way to accelerate such code is to implement the \emph{Single Instruction, Multiple Threads} (SIMT) principle using tools derived from \emph{Message Passing Interface} (MPI) or \emph{Open Multi-Processing} (OpenMP) libraries. Tools in Python related to this approach are presented in Section \ref{sec:distributed_shared}.
%We will first address the seminal parallelization approaches.

\subsection{Distributed memory and shared Memory approaches} \label{sec:distributed_shared}

%As DS and ML algorithms often feature loops, a straightforward path is to try to parallelize these loops.
%For long, parallelization of programs has mainly been performed using two tools: Message Passing Interface (MPI) and Open Multi-Processing (OpenMP). 

MPI works with a distributed memory model, potentially exploiting a distributed network of machines. MPI is a message passing standard that defines the syntax and semantic of library routines to develop parallel applications. With MPI, computers running a parallel program can exchange messages. MPI was originally designed to develop programs in the languages C, C++, and Fortran. Nonetheless, some Python libraries offer the same bindings for MPI, e.g., MPI4Py \cite{DALCIN20051108} and PyPar \cite{pypar2022}.

Conversely, OpenMP works on a shared memory model for multi-core CPUs using program directives.
The OpenMP standard \cite{openmp1998} provides a set of code annotations and instructions for the compiler and a runtime library that extends Fortran and C/C++ languages to express shared memory parallelism. OpenMP is based on compiler directives, thus less intrusive in the code than MPI, i.e., not requiring a strong refactoring of the existing code base. Based on those directives it allows the compiler to parallelize chunks of code whose instructions can be shared among the processors. OpenMP is supported by the most common compilers such as Clang, LLVM, and GCC. It supports loop-level, nested, and task parallelism. Commonly, annotations or directives of the OpenMP API are used in loops. 
OpenMP has two main related implementations in Python; one of the most famous is Pymp \cite{pymp}. It is a library that proposes a special language construction to behave like OpenMP. It relies on the system Fork mechanism instead of threads to make parallel computation. It tries to reduce its footprint by referencing memory and not copying everything in the forked process.
The second one is PyOMP \cite{PyOMP2021}, which is based on Numba and offers a set of constructs similar to the OpenMP API. Nevertheless, the compilation pipeline for Python is a bit more complex: PyOMP uses Numba to generate code in LLVM, then machine code to be able to run it.

%TODO -> be MORE analytics on the tools like the point below
%Difficulties to implement / requieres training and knowledge.
%Maintenance /updating
%

% only subpart of task-based, rest in Section 3
\subsection{Task-based approaches} \label{sec:general_taskbased}

% Task based with library constructs and decorators
%Alternatively, code decorators may be used by some task-based distributed computing libraries.

Task-based frameworks are more recent tools to parallelize and distribute heavy computation. While they are generally powerful, their usage may imply heavy code refactoring. More complex framework, that impose a specific development approach are detailed in Section \ref{sec:distributed}. In this section, we cover libraries which rely on such frameworks, but aim at parallelizing an existing piece of code with minimal impact using decorators.
A decorator is an instruction set before the definition of a function. A decorator indicates that a function (associated to the decorator) must transform a user function (the decorated function) and extend the behaviour of the latter function without explicitly making modifications.

Decorators are used to express parallelism, by indicating that these functions are going to be treated as tasks. The decorated code is analyzed and converted (if applicable) into a suitable version for parallelization. Falling into this category we found PyCOMPSs \cite{Tejedor2017}, Pygion \cite{slaughter2019pygion} and Pykokkos \cite{AlAwarETAL21PyKokkos}, wrappers for COMPSs \cite{Tejedor2008}, Legion \cite{bauer2014legion} and Kokkos \cite{Trott2022}, respectively.
PyCOMPSs and Pygion share some similarities. Both libraries build a task dependency graph and perform analysis to define the order of task execution and the parallelism that can be achieved. Decorators are also similar, as PyCOMPSs and Pygion both use \emph{@task}. On the other hand, Pykokkos translates Python code into the Kokkos API written in C++ and has more decorators to implement its programming model. In PyKokkos for example, functions can be decorated with \emph{@pk.workunit}. These functions can run in parallel by passing them as argument to the function \emph{parallel\_for}. PyKoKKos also has support for using GPUs with CUDA.

In Jug \cite{coelho2017jug}, a task is defined as a Python function, and its arguments take values or outputs of another task. Using the \emph{@taskgenerator} decorator, Jug performs an analysis on a task dependency graph to define the execution order and parallelization of the tasks. Parallelization is achieved by running more than one Jug process for distributing the tasks and using synchronization to get a result. As it is developed with Python, libraries such as Numpy and Scikit learn are compatible with Jug. 

Pydron is a library to parallelize sequential Python code through decorators \cite{Muller2014}. Pydron targets multi-core, clusters, or cloud platforms. First, it translates the decorated functions in Python into an intermediate representation with a data-flow graph structure. The graph is analyzed by a scheduler which defines the order tasks are going to run by putting them in a queue, some tasks being scheduled to run in parallel. When a task is finished, the scheduler must be informed and based on the available information it changes the execution graph. The tasks are distributed to be executed on worker nodes. There is a distribution system in charge of managing the hardware resources, commonly a Python interpreter is launched per CPU core, each in charge to execute a given task.

\subsection{Program transformation and compilation}
\label{sec:programtransformation}

Besides annotations, directives, and decorators for parallelization mentioned in previous sections, program transformation and compilation is another straightforward way to obtain better execution performance for an existing codebase with minimal work overhead for the practitioner. These approaches rely on code analysis that can be either static (source code) or dynamic (based on traced execution) before proposing a transformation of the code into a target language to obtain a better performance in their execution. As such, they can provide performance improvement in a general programming context.

The prominent approach we have found is to guide or give hints to the transformation tool, often a compiler, regarding specific sections of the code that should be optimized. These hints are expressed by the user by typing variables or adding decorators. Additionally,, we review transformation tools which claim full automation, which means input code is passed as it is then.

\subsubsection{Semi-automatic approaches}
Cython, bundled with most Python distributions, is a language extension that serves as a superset of Python. It acts as a compiler, transforming high-level Cython code into highly efficient C code.
One of the key advantages of Cython is its ability to write C extensions, seamlessly integrating them into Python programs for improved performance\footnote{\url{https://cython.org/}}. By translating Cython code into optimized C/C++ code and compiling it as Python extension modules, it unlocks the potential for faster execution compared to pure Python. Remarkably, the majority of Python code can be compiled by Cython without modifications, making it a straightforward tool for optimizing Python applications.
To further enhance performance, it is crucial to add static type declarations to Cython code. These declarations enable the Cython compiler to generate simpler and faster C code. Additionally, Cython provides automatic conversions between Python objects and basic numeric and string types. While Python handles memory allocation dynamically, Cython allows manual memory management similar to traditional C code.

Cython also provides parallelism mechanisms through the module \emph{cython.parallel} using OpenMP as back-end \cite{openmp1998}. To use the parallel module, the GIL must be released. When the GIL is released, Python objects cannot be manipulated. Therefore, a function that deals with Python objects cannot be directly invoked with parallel attributes: the data must be converted into Cython typed variables or memory views.  Good candidates for Cython implementation are general mathematical operations, array operations, and loops. By just using static typing and replacing Python math operations, obtaining a speed up with Cython is highly probable, even if maximal gains require fairly good development skills.
If not well exploited, the performance gain will only be marginal. Moreover, it requires a manual detection of the code parts that could really benefit from Cython, it will depend on the ability of a programmer to use profilers to find out the bottlenecks of the execution of a program.

JIT stands for \emph{Just-in-Time} compilation, a technique employed by some programming languages and runtime environments to enhance code execution performance. Python is an interpreted language, which means that code is executed directly by the interpreter without prior compilation. While this approach offers flexibility and dynamic features, it can be slower compared to compiled languages. However, JIT compilation allows the interpreter to dynamically compile specific code sections into machine code just before execution. This runtime process optimizes the code for improved performance. Some alternative interpreters to CPython use JIT compilation to optimize their performance.  As we left alternative interpreters out of the main scope of our paper, please refer to Section \ref{sec:interpreters} for more information in this area.

A highly popular JIT compiler for Python is Numba \cite{numba2015}. Numba provides compilation of Python code for a faster execution. The user must use decorators to indicate code parts that should be improved by the compiler. A common function decorator in Numba is \emph{@jit} and has the following parameters: \emph{nopython}, \emph{parallel}, and \emph{fastmath}. If \emph{nopython} is set to true, the JIT compiler would compile the decorated function so it will try to run without the involvement of the Python interpreter. The \emph{parallel} flag enables Numba with a transformation pass that will attempt to automatically parallelize and/or perform other optimizations on the function or some parts of it. The \emph{fastmath} flag relaxes some numerical rigor to gain additional performance and enables possible fast-math optimizations. By executing the code, the Numba JIT would attempt to apply the improvements we indicated with the decorators and their parameters.

%\begin{lstlisting}[float, style=pythonstyle, caption={Numba simple example}, label={numbaex}]
%from numba import jit # Numba import
%import numpy as np
%
%x = np.arange(100)
%
%@jit(nopython=True, parallel=True, fastmath=True) #Numba decorator and parameters
%def do_something(a): 
%t = 0.0
%for i in range(a.shape[0]):
%	t += np.sin(a[i])
%return a + t
%
%print(do_something(x))
%\end{lstlisting}

The Numba compiler translates Python code into an intermediate representation, then it is translated to LLVM to finally emit machine code. The generated machine code is close in terms of performance to a traditional compiled language. Similarly to Cython, Numba only supports a subset of the Python language and some specific libraries like Numpy.
Numba can convert a sequential code to be executed in parallel by multiple cores and in very limited cases to be executed in a GPU. 
Numba can also be used as a bridge to develop programs in Python to run in the GPU. It offers support for CUDA (Nvidia hardware), ROCm (AMD) and HSA (AMD and ARM). A big difference is that there are no automatic attempts to parallelize the code then. Instead, the user must re-factor the code to a style similar to C with CUDA. Numba can compile a restricted subset of Python code into CUDA kernels and device functions, HSA kernels, and ROCm device functions. In GPU programming, a kernel is a GPU function launched by the host (CPU and its memory) and executed in parallel on the device (GPU and its memory). A device function is a GPU function executed on the device which can only be called from the device.

Designed within the context of astrophysical applications, Hope specializes in numerical computations.
Hope is a JIT compiler that uses the decorator \emph{@hope.jit} with the function to be translated. The decorated functions are parsed into a Python Abstract Syntax Tree (AST). The Python AST is converted into a Hope AST. Several optimizations may be applied to the Hope AST such as simplification of expressions, factorizing out subexpressions, and replacing the \emph{pow} function for integer exponents. From the Hope AST, C++ code is generated and compiled into a shared library (\emph{.so} file on Linux systems). The shared library is added to the cache, loaded, and executed. Hope validates the name of the functions and the types of the passed arguments and tries to match to what it has on the cache, if not found then the whole compilation process starts over. The data types used in the functions are inferred by static analysis of code, the AST, and the runtime analysis. The simplification of expressions and common sub-expression elimination is performed with the SymPy library \cite{Sympy2022}. 

Autoparallel \cite{Ramon-Cortes2020} is a compiler for Python code to transform nested loops from sequential to parallel execution in a distributed computing infrastructure. It requires that the user adds a decorator on identified functions that contain nested loops. Autoparallel relies on PyCOMPSs \cite{Tejedor2017} and PLUTO \cite{Bondhugula07pluto:a}; PyCOMPSs is a task-based programming model to develop applications with Python decorators (reviewed in Section \ref{sec:general_taskbased}), whereas PLUTO is a parallelization tool that automatically transforms affine loops using the polyhedral model \cite{bastoul_code_2004}. Autoparallel analyses code decorated with \emph{@parallel} and for each affine nested loop that finds creates a \emph{Scop} object. The Scop object is then parallelized by adding OpenMP-like decorators to the loops. Then, it converts the code into task format through PyCOMPSs by adding tasks configurations and data synchronizations. Finally, each nested loop is replaced by the generated code to be executed by PyCOMPSs in a distributed computing platform.

\subsubsection{Automatic approaches} \label{sec:automatic}
Transforming software in view to maximize performance is difficult to perform fully automatically. Code translation and transpilation focus on analyzing the structure of the code and apply transformation patterns as means to circumvent the absence of supervision.

Due to the nature of Python as an interpreted language, an increase of performance can be obtained by just porting a Python program into a compiled language. Nonetheless, doing it manually is a cumbersome task. Therefore, some specialized libraries perform transpilation by translating Python code into a compiled language (most often C++). 
%Following this principle, the following libraries translate Python code into C++: Hope \cite{AKERET20151}, Shed Skin \cite{shedskin2022}, Nuitka \cite{nuitka2022}, and Pythran \cite{Guelton2015}.

Shed Skin uses static analysis by checking implicit types of variables. Therefore, Shed Skin requires that all variables are implicitly typed. In other words, they must only have one assignment, and multiple assignments of different types to the same variable is not supported. To use Shed Skin, a command must be used in a terminal and the file containing Python code is passed as an argument. The Shed Skin compiler generates the translated code in C++, a header file, and a make script to compile it. Moreover, a module can be compiled and invoked from another Python script.

Nuitka translates CPython instructions into a C++ program. Compiled code generated by Nuitka is executed along with the Python interpreter for the part that cannot be compiled. This means that compatibility with other libraries is supported while using Nuitka. No code modification is required. To use Nuitka the code must be compiled using the console through Nuitka commands along with the Python code filename. The code and executable files are generated and can be invoked directly or as stand-alone libraries. 

Pythran converts Python code into C++ code. However, it goes beyond pure translation and performs code analysis and optimizations. Pythran receives as an input a Python module meant to be converted into a shared library. On the front-end of Pythran, the Python module is converted into a Python AST. Then, the Python AST is converted into a Pythran internal representation (IR) which is a subset of the Python AST. During this conversion, code analysis steps and different transformations and optimizations are performed, aimed at generating a faster version of the code. Additionally, variable types may be inferred by static analysis. The back-end of Pythran turns Pythran IR into parametrized C++ code. Then, Pythran instantiates and compiles the generated code to build a native module. Pythran is compatible with Numpy expressions and applies optimizations such as expression templates, loop vectorization, and loop parallelization through OpenMP.

Transpyle \cite{Bysiek2018TowardsPH} relies on transpilation to accelerate Python performance. The originality of this approach is to support multiple languages also as input, e.g., reusing a legacy optimized loop written in Fortran and integrate it in the transpiled Python code. Moreover, with the use of Python as the intermediate representation for compiling code from and into target languages (e.g., Fortran), it helps the Python developer to understand the complete process. It also works in a semi-automated mode with Python annotations, possibly guiding the compiler for better improvements (e.g., loop unrolling and vectorization).

%\begin{lstlisting}[float, style=pythonstyle, caption={Transpyle annotation example for loop unrolling}, label={transpyle}]
%@transpyle.unroll('i', 4)
%def elementwise_add(arr1, arr2):
%    assert len(arr1) == len(arr2)
%    arr3 = np.array((arr1.size,), dtype=float)
%    for i in range(0, len(arr1)):
%        arr3[i] = arr1[i] + arr2[i]
%    return arr3
%\end{lstlisting}

ALPyNA \cite{alpyna2019} is a program transformation tool for Python which uses static and dynamic analysis of nested loops and generates CUDA kernels for GPU execution. 
The input code must contain vanilla Python code and optionally Numpy instructions.
Currently, basic subscripting of single or multi-dimensional arrays is supported, i.e., no slicing or sequence indexing. ALPyNA performs analysis mostly on nested loops, where a performance bottleneck is more probable to occur. Other Python instructions are ignored and are executed by the Python interpreter. 
After static analysis, if loop bounds and data dependencies can be determined, ALPyNA generates untyped GPU kernels. 
Otherwise, loops are marked for analysis at runtime. For runtime analysis (and execution) the ALPyNA execution object must be used (obtained by the function that performs static analysis) to invoke the original functions. If possible, loop bounds and data dependencies are determined at runtime and GPU kernels are generated on the fly. ALPyNA relies on Numba to finalize and compile the GPU kernels.

Pyjion is a JIT compiler designed to improve the performance of Python by converting CPython bytecode into machine code \cite{pyjion2023}. In the absence of Pyjion, CPython relies on a master evaluation loop known as the frame evaluation loop to sequentially execute opcodes (individual instructions within the bytecode). However, Pyjion's compiler consists of three key stages. First, it constructs a \emph{stack table} that maps abstract types to each opcode position. Then, it compiles CPython opcodes into Common Intermediate Language (CIL) opcodes. Finally, it emits these CIL opcodes to the .NET Execution Engine (EE) compiler, which converts them into native machine code or assembly. Overall, Pyjion enhances the execution speed of CPython by leveraging various optimizations in a JIT compilation context.

%Pyjion does not currently support with blocks.
%Pyjion does not currently support async await (YIELD_FROM) statements.

Pyston \cite{pyston2022} is an alternative Python interpreter that includes a JIT step as well as many performance optimizations for the execution of Python programs (discussed specifically in Section \ref{sec:interpreters}).  Pyston-lite is a derivative of Pyston that just retains the JIT step, but with a focus on easier installation and setup. While Pyston aims to achieve the highest performance possible, Pyston-lite may not match the same level of performance as the full implementation of Pyston. However, it still offers improved speed compared to using CPython as the interpreter: the authors claim 10\% acceleration on macrobenchmarks \cite{pyston2022}. It's important to note that Pyston is available for Python version 3.8.12, while Pyston-lite supports a wider range of Python versions: 3.7 to 3.10. 

Let us note that in general, the performance of most tools in this section has been estimated by contrast to vanilla Python. While this suits the use case scenario motivating the section, in general performance does not increase monotonically by combining the usage of multiple tools. For example, while Pyston-lite significantly improves the performance of vanilla Python, we observed that using it in combination with Numba parallelization tended to degrade the performance of Numba obtained with CPython (e.g. the execution time of matrix multiplication code has more than doubled in some of our experiments).

\section{Accelerating numerical libraries usage}
\label{sec:libraries}
In this scenario, an algorithm would have already been implemented by the data scientist, but contrasting with the previous section, it would not rely only on vanilla Python, and would also use Python numerical libraries. Indeed, it would have been recognized that the problem depends mostly on standard data structures such as float matrices, thus aiming at benefiting from associated out-of-the-box primitives (e.g., matrix decomposition algorithms). In this section, we focus on means to provide faster execution of such numerical libraries or APIs. 

For the sake of clarity and legibility, we focus on the three main libraries used in DS to facilitate and accelerate the development of single-threaded numerical computation code: Numpy \cite{harris_array_2020}, Pandas \cite{pandas}, and Scikit-learn \cite{scikit-learn}.  These libraries are always among top results when looking for cornerstone libraries to carry out statistical and numerical computations needed for the practice of DS and ML. They are ranked 20$^\text{th}$, 27$^\text{th}$, and 103$^\text{rd}$ in terms of monthly downloads on PyPI (Python Package Index) in 03/2023, respectively.
It is worth mentioning that other libraries are widely used in DS and ML. However, they are tied to secondary tasks such as preprocessing (e.g., NLTK, ranked 334$^\text{th}$) or visualization and plotting (e.g., matplotlib, ranked 114$^\text{th}$). As this survey focuses on accelerating DS code, we do not directly cover these libraries in this section. Statsmodels is also a frequently mentioned library, but we did not retain it, as it is a bit further down in the PyPI download ranking (327$^\text{th}$).

Besides approaches covered in other sections (e.g., compilation, transformation), in the context of these libraries we mainly found solutions implementing an API with the same signature (same inputs and same outputs) as the original but proposing better performance. We refer to these as \emph{drop-in} libraries. The execution of those drop-ins can be done using multiple CPUs, GPUs, and/or with a more efficient implementation. It may eventually require minor modifications such as data copies and changing function parameters. Ideally, they bear minimal cost to the practitioner in terms of development overhead.
In this section, we will review the three identified libraries and their performance enhanced counterparts.

% Python drop-in package amending keywords

\subsection{Numpy} \label{sec:numpy}
It is one of the most used Python libraries, as it provides a multi-dimensional array format central to many other libraries. It also includes a set of routines for manipulating arrays with different operations, e.g., mathematical primitives, shape manipulation, and sorting. 
Numpy exploits BLAS and LAPACK and is therefore much faster than vanilla Python code. However, it under-utilizes parallel computer architectures. Several examples of Numpy drop-in libraries attempt to circumvent this issue.

\subsubsection{Legacy drop-in}

Distarray \cite{distarray2022} is a drop-in library for Numpy, which distributes the execution of Numpy operations across multi-core CPUs, clusters, or supercomputers. It depends in IPython.parallel \cite{ipyparallel2022} (surveyed in Section \ref{sec:distributed}) and MPI for setting up a cluster. Closely related is DistNumPy \cite{Kristensen2010} which implements parallel Numpy operations by also using MPI underneath. DistNumPy was deprecated and moved to Bohrium which is in active development. 

Bohrium \cite{kristensen2013bohrium} is a runtime that maps Numpy array operations (universal functions, also known as \emph{ufuncs}) onto different hardware platforms such as multi-core CPUs, GPUs, and clusters. To use Bohrium the user must either replace the Numpy library import with the bohrium library or launch a script with the command \emph{python -m bohrium myscript.py}. Bohrium uses different techniques to speed up computations. For example, Bohrium supports lazy evaluation, this means that Numpy operations are regrouped for evaluation until a non-Numpy operation is found. Bohrium fully supports Numpy views, therefore no data copies are done when slicing arrays. When certain conditions are met, array operations are fused into a single kernel that is compiled and executed. Data copies between main memory and GPU memory are done only when the data is accessed through Python or a Python C extension.

Bohrium is built with components that communicate by exchanging a \emph{vector bytecode} (an intermediate representation corresponding to the Numpy array operations). The instruction (original code) is passed to a \emph{Bridge} component which generates the vector bytecode. This bytecode is passed to a \emph{Vector Engine Manager} component which manages data location, ownership of arrays, and the distribution of jobs between vector engines. The Vector Engine component is an architecture-specific implementation to execute the bytecode. Non-Numpy or unsupported operations fall back into the regular CPython interpreter. While setting up Bohrium is fairly straightforward, experimentally we found that its value as a drop-in for Numpy remains limited: for example, the Singular Value Decomposition, commonly used in the practice of DS, even causes a crash instead of gracefully falling back to Numpy\footnote{Tested with the latest version on PyPI on 03/2023}. Also, we did not find cases where using the GPU as a backend did not lead to strong performance degradation.

% PBR: I completely removed this block, as by reading it I have absolutely no idea in what npbackend is supposed to do. Unless a clearer and unambiguous description can be written, I am not in favor of retaining this, especially as the tool is deprecated.

%Although not a direct implementation of Numpy, npbackend \cite{kristensen2014separating} is a library to facilitate the implementation of different Python accelerators as back-end for Numpy operations. npbackend works as a middleware between python Numpy operations and a back-end. For the end-user it looks like as a drop-in replacement, requiring only that the user installs the package and modify the library import. It is mainly composed by two interfaces: user and back-end. The user interface main task is to move from regular NumPy code to a unified NumPy back-end. The back-end interface is in charge of isolating the operation-specification from the implementation. Numpy instructions are translated into the specific calls according to the back-end. npbackend uses its own array structure (\textit{npbackend-array}) which inherits from Numpy-array. With this approach all Numpy-operations are accessible and a subset of the operations are replaced (overloaded) with an specific implementation. npbackend does not rewrite any Numpy operations but merely serves as a middleware between the Numpy API and execution tools. Additionally, it implements optimizations such as memory allocation reuse. It is noteworthy that an implementation of npbackend was built with Python 2.7 and it is integrated with Bohrium.

D2O \cite{steininger2016d2o} is a middleware between Numpy arrays and a distribution logic. In that sense it is not a drop-in library, but an interface to provide parallel execution of Numpy array operations through the use of a \emph{distributed data object} format. The user can pass a Numpy array as an argument to create a distributed data object, along with options regarding distribution strategy. The distributed data object supports many Numpy instructions such as arithmetic operations, indexing, and slicing.
% PBR: I don't get it - "under the hood" and "user must create" are contradictory to me
D2O relies on MPI4Py to distribute the work (see Section~\ref{sec:generalpurpose}). Therefore, to exploit parallelism with D2O the user must create an MPI job. The number of nodes can be specified on the command to run the Python program. For lower-level instructions the MPI library is accessible for code refactoring. 

%\begin{lstlisting}[float, style=pythonstyle, caption={D2O basic example from \cite{steininger2016d2o}}, label={numexprcode}]
%import numpy as np
%from d2o import distributed_data_object 
	
%a = np.arange(16).reshape((4,4))
%ob = distributed_dat_object(a)
	
%# doing a series of simple arithmetic operations
%(2*obj, obj**3, obj>=5)
%\end{lstlisting}

% PBR: this sentence is not very clear to me - I have the impression that it is not necessary to understand the general message either
%Given that the data is partitioned between nodes, the output of the distributed data objects can either show the local portion or all the data of the global data available.

\subsubsection{GPU acceleration}

Many Numpy functions boil down to the same operation applied to large vectors or matrices, and can thus exploit GPUs acceleration. CuPy~\cite{nishino2017cupy} was designed to cover the API of Numpy as widely and transparently as possible.
CuPy uses the Nvidia CUDA framework and other CUDA libraries for optimization such as cuBLAS, cuDNN, cuSPARSE. Given the differences of memory management between the main memory and GPU memory, for harnessing the library at its best, the user must manually indicate data copies, so that data is available in the GPU memory when CuPy functions are called. However, the process remains straightforward compared to CUDA programming. For example, using CuPy in the context of a semi-supervised learning task allowed to divide baseline computation time by 6 with reasonable implementation efforts \cite{castro_parallelization_2022}. For cases where the available functions are not enough, CuPy supports creating user defined CUDA kernels for two types of operations. One is for element-wise operations where the same operation is applied to all the data. The other operation is for reduction kernels, which folds all elements by a binary operator.

In the line of Numpy drop-in libraries for GPUs there is also PyPacho \cite{pypacho2021} and DelayRepay \cite{Morton2020}. PyPacho is based on PyCUDA and PyOpenCL. Although it is a promising tool, it is not as mature as CuPy and offers less compatibility. On the other hand, DelayRepay is a drop-in library and applies code optimization to accelerate its execution. DelayRepay has a delayed execution of Numpy operations because it analyzes them and tries to fuse them before execution. When a Numpy operation is found, it checks if its output is the input of another Numpy operation. If the rule is fulfilled, the operations are fused and the AST is modified. The Numpy operations are fused until a non-Numpy operation is found. When a non-Numpy operation is found, the fused AST node is compiled into a GPU kernel and executed in the GPU. This is a main difference compared to CuPy which executes each operation individually.

Although not a drop-in library for Numpy, PyViennaCL \cite{pyviennacl2022} provides a set of equivalent operations to be executed in multi-core CPUs and GPUs. PyViennaCL is a wrapper for ViennaCL (written in C++) which is a linear algebra library and numerical computation to execute on heterogeneous devices. To use PyViennaCL, the user must import the library and use the constructs provided by the library. Similarly to DelayRepay, PyViennaCL uses delayed execution. Arithmetic operations are represented by a binary tree and are computed only when the result of the computation is necessary.

\subsubsection{Compilation-based}

The JAX \cite{jax2018github} library provides composable transformations of Python programs based on Numpy. All JAX operations are implemented using the Accelerated Linear Algebra compiler (XLA) \cite{xla2020}. JAX provides a set of equivalent functions to Numpy. Therefore, it can be used as a drop-in library for Numpy. Besides performance improvement based on vectorization and parallelization, JAX provides JIT compilation into GPU or TPU using the \emph{jit} function. Another functionality is the evaluation of numerical expressions and generating derivatives (e.g., automatic differentiation by passing functions to the function \emph{grad}), as commonly used by gradient methods for training neural networks. Another important functionality in JAX is \emph{vmap} which is a mapping function to vectorize operations. The \emph{jit} function can be applied to \emph{grad} and \emph{vmap} to obtain better performance results.

An option specialized in speeding up numerical expressions written in Numpy is NumExpr \cite{numexpr2022}. This library is compatible with a subset of Numpy operations. To use it, expressions are passed as a string to the library function \emph{evaluate}. The expression is compiled into an object that contains the representation of the expression and the types of the arrays. To validate the expression, first it is compiled by the Python \emph{compile} function, the expression is evaluated, and the parse tree is built. The parse tree is compiled into bytecode, and a virtual machine uses \emph{vector registers}, each with the same fixed size. Arrays are handled as chunks, these chunks are distributed among the CPUs to parallelize Numpy operations. This approach has a better usage of cache memory and can reduce memory access, especially with large arrays.

%\begin{lstlisting}[float, style=pythonstyle, caption={Numexpr basic example}, label={numexprcode}]
%import numpy as np
%import numexpr as ne
%
%x = np.arange(1e6)
%y = np.arange(1e6)
%
%ne.evaluate("x * y + 10")
%\end{lstlisting}

In this inventory we may also mention work surveyed in the previous section~\ref{sec:programtransformation} like AlPyNa, Pythran and Numba. These tools have general applicability for Python performance improvement, but also provide performance improvements specific to Numpy.

\subsection{Pandas}
Pandas is a highly popular Python library for data analysis and manipulation. Its data frame format is widely used in DS, as it notably allows to handle heterogeneous data, time series and query-based manipulation, to name a few features. A data frame is a two-dimensional data structure that contains labelled axes: row and columns. It is the primary data structure used in data analysis tools. Nonetheless, Pandas operations usually only use one core at a time when doing computations. Thus, multi-core CPU and GPU oriented drop-in libraries have emerged to accelerate Pandas-like operations. 

Vaex is a library that contains a set of packages meant to optimize memory usage when managing large datasets \cite{breddels2018vaex}. Vaex-core is a drop-in library for Pandas-like operations on data frames. Similarly to Bohrium \cite{kristensen2013bohrium} with Numpy, most operations on Vaex are lazily evaluated, they are computed only when needed. This reduces the amount of memory required compared to other similar libraries. Vaex also works with small chunks on data on the RAM, therefore, it can work with datasets larger than the typical RAM of a computer. It works best with files in HDF5, Apache Arrow, and Apache Parquet formats.

A multi-core CPU drop-in implementation of Pandas is Modin \cite{petersohn2020towards}. Modin can perform in a single node locally (multi-core CPUs) or in a cluster environment. Modin is based on a custom version of the Pandas data frame, and treats operations on the data frames as user queries which are compiled. Having a similar design as relational databases which work with relational algebra, Modin is designed to work with a custom data frame algebra, aiming at simplifying and optimizing operations on a data frame. The Pandas-like API instructions are translated into data frame algebra with optimizations if possible.  Then, the optimized query is passed to a subsystem called Modin Dataframe which works as a middle layer between the query compiler and the actual execution back-end. A data frame can be partitioned by columns, rows, or by blocks depending on the operation required and the size of the data, enabling work distribution. In local mode the number of partitions is by default equal to the number of available CPU cores. The Modin dataframe subsystem passes the data to the execution layer where different execution engines can be used such as Dask \cite{rocklin2015dask} or Ray \cite{moritz2018ray} (see Section \ref{sec:distributed} for an introduction of the latter) which are in charge of the actual execution of computations on partitioned data in a task-based approach. The installation with the Ray backend is straightforward, with 40 times acceleration for a column concatenation benchmark, but mitigated results otherwise. Also, moderate code adaptation will generally be needed, as Modin return formats often slightly differ from their pandas counterparts.

cuDF \cite{cudf2022} is a Pandas drop-in library that runs on the GPU. It is used for manipulating data with the GPU for data science pipelines. cuDF is a building block of RAPIDS, a platform to execute ML and DS tasks in GPUs (see Section \ref{sec:scikitlearn}). Dataframes can be created, read from files, converted from Pandas dataframes and CuPy arrays. 
Some tools, though not drop-in libraries for Pandas as such, bear high similarity with Pandas, to such an extent that minor refactoring to the code can be used for the same purpose. Following this approach, we found Datatable \cite{datatable2022} and Polars \cite{polars2022}. Datatable is implemented in C++ and uses multithreading for certain operations to speed up processes. Polars lazily evaluates queries to generate a query plan and optimizes it so it can run faster and reduce the memory usage, possibly exploiting parallelism. Both libraries can also easily export to and from Numpy and Pandas formats.

\subsection{Scikit-learn} \label{sec:scikitlearn}
% scipy: 74th
Scipy \cite{2020SciPy-NMeth} reuses the array format defined by Numpy, but aims at a more comprehensive coverage of general purpose mathematical and statistical concepts, such as linear algebra, statistical tests, signal and image processing. Scikit-learn builds upon Numpy and Scipy by implementing many models from the ML literature, such as regression, classification, and clustering models. Most models implement \emph{fit} and \emph{predict} functions, providing a unified API for the library.
%if we maintain this sub-section we will require more references

\subsubsection{dislib}

Dislib \cite{dislib2019} is a ML library for Python to be executed in high-performance computing clusters. Dislib is built on top of PyCOMPSs \cite{Tejedor2017} (a task-based parallelization library presented in Section \ref{sec:distributed} in the context of the first scenario) and exposes two main components to the developers: 1) an interface for distributed data handling and 2) an estimator-based API. The data handling interface provides an abstraction to handle data as a dataset which can be divided in multiple subsets to be distributed and handled in parallel. Datasets can be given as Numpy arrays for dense data and Scipy Compressed Sparse Row matrices for sparse arrays. Its wrapping Dataset format is the input for the ML models.

The estimator API provides a set of ML models with a similar syntax as Scikit-learn. An estimator is an abstraction of a ML model and typically implements two characteristic methods in Scikit-learn: fit and predict. To summarize, data is loaded into the Dataset format. An instance of an estimator object (representing the ML model) is created, and the fit function is invoked with its parameters. The estimator object is used to retrieve information of the trained model and generate predictions. Dislib authors report that they outperform MLlib \cite{meng2016mllib} for a k-means clustering task as the sample size and the number of computation nodes used grows very large.

\subsubsection{cuML}

RAPIDS \cite{rapids2022} is a set of libraries for data manipulation and machine learning developed on top of the CUDA language, and thus aimed at the execution of DS pipelines in GPUs. In this set of libraries, cuML is strongly related to Scikit-learn.
As CuPy aims at covering most of the Numpy API, cuML was created with the target to cover as much of the Scikit-learn API as transparently as possible. Similarly, as Scikit-learn is built on top of the Numpy and Pandas formats, cuML exploits the CuPy array and cuDF dataframe formats, respectively. Most of its API can also be executed in a distributed environment using Dask, a task-based framework relevant to the scenario presented in Section \ref{sec:frameworks}. The cuML developers report that GPU implementations can run up to 50 times faster than their scikit-learn CPU-based counterpart\footnote{https://github.com/rapidsai/cuml}.

\subsubsection{MLlib}

MLlib \cite{meng2016mllib} is a ML library part of the Spark system. It is similar to Scikit-learn with a set of ML models and data processing instructions. Built on top of Spark it thus comes with the Spark installation and a Python API to use it. The implementation of algorithms is parallelized so that large data processing jobs exploit data distributed on Hadoop clusters.

\section{Structuring Frameworks} \label{sec:frameworks} % not sure about this title
In this section, we consider high-performance libraries and frameworks which impose a specific way of thinking and programming to the practitioner and are thus preferably used right when implementation starts.

% First, focus on what is often referred to deep learning frameworks - define in general terms of computation graphs and symbolic computing as a specific problem, yielding embarrassingly parallel problems, with solutions benefiting from GPU hardware

\subsection{Deep Learning frameworks}

% general terms on deep learning libraries, chained to TF
Many models used in DS can be formalized as Directed Acyclic Graphs (DAG), e.g., Bayesian networks, probabilistic mixture models, and most notably, neural networks. A range of Python libraries, commonly referred to as \emph{deep learning frameworks}, comes with specialized support and useful abstractions to practitioners needing to put this kind of models in action. Computations underlying DAGs are typically embarrassingly parallel: benefiting from high-performance computation devices such as multi-core CPUs or GPUs is therefore an implicit requirement of these libraries. Technically, they are symbolic mathematical libraries which allow to define arbitrary computational DAGs along which data is transformed. However, their \emph{deep learning} label is often well deserved, as they provide many facilities specifically oriented towards neural networks, such as automatic gradients and back-propagation at DAG nodes, enabling fitting model parameters to input data. At runtime, the computational graphs and all functions which operate on them (e.g., custom loss functions and gradient optimizers) are compiled and loaded to the GPU. The training procedure then triggers kernel execution on the GPU. While these frameworks can also be run on multiple cores of a CPU when no compatible GPU is available, taking effective advantage of the GPU generally yields very significant speedup. For example, the computation time of a piece of Tensorflow code heavily relying on the CPU can be divided by 10 if correctly using the GPU API \cite{castro_parallelization_2022}.

Tensorflow \cite{tensorflow2015-whitepaper} is the most prominent in this range of tools. 
Besides offering a wide range of ready-to-use model architectures (sometimes even along pre-trained model weights), Tensorflow defines a comprehensive API to program custom components then compiled and loaded on the GPU, such as model structures, loss functions or optimizers. As this code is meant to be loaded on the GPU, although it uses the Python syntax, it cannot be mixed with regular Python instructions, which causes additional implementation effort. In Tensorflow, the computational DAG is defined statically, so that its compilation and execution yields maximum performance at runtime. The explicit definition of the computational graph and its asynchronous execution on the GPU yields constructs which tend to diverge from Python standards. Mastering Tensorflow therefore takes some time and practice.

% PyTorch - pros and cons with TF

Torch is another deep learning framework, developed by Meta with the similar aim to support neural network model training. However, it is based on the Lua language, which is limiting its popularity. PyTorch \cite{pytorch} is the port of Torch to Python, motivated by the will to keep its API and basic principles.
PyTorch came to the market after Tensorflow, but has gained momentum and is catching up in terms of popularity (8M monthly downloads vs 15M for Tensorflow according to PyPI statistics\footnote{\url{https://pypistats.org} was accessed on 28/10/2022}). Good documentation facilitates its adoption by newcomers, and it offers many ready-to-use model architectures and pre-trained parameters. PyTorch has built-in high-level APIs, which are delegated to Keras in the case of Tensorflow. Pure Tensorflow requires significant non-standard boilerplate code development in comparison.

In Tensorflow, the computational graph is defined and compiled statically, and placeholder data is replaced at runtime. PyTorch offers more control at runtime, e.g., allowing to modify execution nodes at runtime in ways forbidden by Tensorflow, facilitating the implementation of sophisticated training loops. Language constructs are closer to Python standards, with object-oriented constructs meant to be familiar to experienced programmers. Overall, its APIs are less rigid, but this comes at the cost of more code to write, and generally slightly longer execution time for equivalent tasks.

%Rephrase
This distinction between static and dynamic computational graphs has other consequences, first in the way Tensorflow and PyTorch handle variable-sized input data. Due to the static computation graph approach, doing so is difficult with Tensorflow. The Tensorflow Fold tier library offered limited support, but it is no longer maintained. In contrast, this is built-in in PyTorch. 

Debugging PyTorch is also straightforward, while it is more difficult with Tensorflow due to the static graph definition. In the latter case, this requires mastering a specific debugging tool, \emph{tfdbg}. 
To compensate, Tensorflow comes with Tensorboard, which packages visualization and monitoring tools. In PyTorch, to come up with equivalent features, custom graphs have to be built using e.g., matplotlib, or an interactive plotting library such as Dash.
More facilities exist for distributed training in Tensorflow, as well as deployment to production servers, and embedding in limited resource devices such as mobile and Raspberry Pi using Tensorflow Lite. Finally, Tensorflow supports several languages beyond Python (including C++ and Java), while PyTorch focuses on Python. 
%TODO checks Torch Lua... PyTorch can be used from Java...

% other DL libraries
Theano \cite{theano} offers very similar features to Tensorflow and PyTorch, primarily aimed at defining and training neural network structures. It has been around since 2007, but its development has been stopped - the latest release dates back to 2020. It has been forked and repurposed to Aesara \cite{aesara}, the latter being aimed at optimizing and evaluating mathematical expressions involving numerical arrays and symbolic inputs. Aesara has therefore more general applicability, comparable to numerical libraries such as Numpy (see Section \ref{sec:numpy}), but it involves computational graphs, and therefore cannot be included in regular Python projects in a straightforward way.

Frameworks such as Tensorflow and Theano require asynchronous thinking, and significant overheads in boilerplate code development. Keras \cite{chollet2015keras} came as a high-level library meant to accelerate experimentation with such deep learning frameworks. It abstracts them with convenient input-output primitives and a simpler training API, notably, while allowing to access and parametrize the backend framework. Overall, it allows the data scientists to program in a more procedural fashion. Up until version 2.3\footnote{Released 7/10/2019}, it supported multiple frameworks: Tensorflow, Theano, and the Microsoft Cognitive Toolkit (formerly known as CNTK). Since version 2.4, only Tensorflow is supported, which means it is now part of the Tensorflow distribution, in practice.

MXNet \cite{mxnet} claims high flexibility and scalability, notably supported and used internally by Amazon. Like Tensorflow, MXNet supports several languages beyond Python (C++, Python, R, Scala, Matlab), when PyTorch focuses on Python. It offers a flexible front-end, with an imperative API meant to be familiar to newcomers, and a symbolic API aimed at maximizing performance. However, it lacks high-level IO primitives compared to PyTorch and Keras, which is detrimental to quick adoption.

% Task-based 
% moved from general-purpose (+ some elements moved back to Section 1)
\subsection{Distributed computation frameworks} \label{sec:distributed}

% Then on more general task-based frameworks, which can sometimes be articulated with above-mentioned frameworks. They have more general applicability, but are similar in that they impose some strong concepts and constructs, pretty much kind of DSLs.
An approach used by multiple Python intensive computation libraries is task-based parallelization, especially when large sets of data are involved. The task-based approach refers to a strategy where the work is divided into multiple tasks, these tasks are handled by a task manager which assigns them to threads that execute them. The execution of a program is a sequence of tasks and in some cases independent tasks can be executed in parallel. Usually, the task-based approach is implemented with a queue of tasks, a thread-pool where threads wait for a task assignment, and some message protocol (i.e., MPI) to communicate data and instructions between tasks and the task manager. Though of general applicability, most libraries in this section impose in depth modifications to an existing codebase and require heavy software setup. This contrasts with task-based approaches reported in Section \ref{sec:general_taskbased}, the complexity of which being scaffolded using decorators.
This makes them a more suitable choice if algorithm implementation has not started yet.

% transition to task-based approaches
Directly relating to deep learning frameworks presented in the previous section, Horovod \cite{sergeev2018horovod} aims at facilitating the usage of distributed resources (i.e., multiple computation nodes, potentially each holding multiple GPUs) by these frameworks. Indeed, deep learning frameworks are sometimes packaged with modules dedicated to distributed training, but, in the case of Tensorflow for example, they are rigid and difficult to set up. Horovod compensates this problem, while offering the support to multiple frameworks (including TensorFlow, Keras, PyTorch, and MXNet). Behind the scenes, Horovod relies on a message passing layer, which can be OpenMPI for example (presented in Section \ref{sec:generalpurpose}). The default is to use Gloo \cite{gloo}, a communication library developed by Meta. The authors claim up to 90\% scaling efficiency, depending on the neural architecture at hand \footnote{https://horovod.readthedocs.io/en/stable/summary\_include.html}.

% Task based but only library constructs/instructions

Some task-based parallel Python libraries we found are wrappers of an already existing library in a different language. This is the case of torcpy \cite{HADJIDOUKAS2020100517} and Charm4py \cite{Galvez2018}. Both libraries are wrappers of their C/C++ counterpart library; torcpy for TORC \cite{Hadjidoukas2012} and Charm4py for Charm++ \cite{kale1993charm++}. In both libraries the parallelism is expressed by using the library instructions and an API lets the programmer orchestrate asynchronous tasks and distributed objects. In torcpy, tasks are executed by launching multiple MPI processes using one or multiple worker threads. It builds upon MPI4Py (see Section \ref{sec:distributed_shared}), and implements and API meant to upscale the logic underlying \emph{multiprocessing} or \emph{concurrent.futures} packages in order to benefit from high-performance clusters. Multiprocessing or concurrent.futures packages and Python built-in packages aiming at overcoming limitations imposed by the GIL, by implementing a parallel \emph{map} function, and allowing the asynchronous execution of functions in multiple parallel Python processes, respectively.
Torcpy claims 90\% computation efficiency on a Monte Carlo molecular simulation tasks with 1,024 compute nodes.
While torcpy is based on Python dictionaries for its data management, in Charm4py multiple distributed objects are executed and coordinated in a unit called processing element. Distributed Python objects are implemented and allow remote method invocation using message passing. To overpass the GIL lock of only one thread, the implementation of Charm4py launches the Python executable in multiple nodes or even multiple times on the same node thanks to this distributed object mechanism. Unfortunately, it does not seem to run any more in recent Python environments\footnote{Tested on 30/03/2023 with Python 3.9}.

% Task based written entirely in Python (no-wrappers)
There are also task-based parallel libraries written mostly or entirely in Python, such as Scalable Concurrent Operations in Python (SCOOP) \cite{Geoffroy2014}, Parallel Python \cite{parallelpython2022}, Celery \cite{Celery2022}, and Playdoh \cite{rossant2013playdoh}. 
SCOOP is similar in spirit to MapReduce frameworks \cite{dean2008}, as its API revolves around \emph{map} and \emph{reduce} functions. Asynchronicity and parallelization is enabled by a custom implementation of the built-in Python \emph{futures} class. Specifically, their workers act as independent elements that interact with a broker to mediate their communications. The documentation gives extensive instructions to facilitate the deployment on high-performance clusters.
Parallel Python also relies on its own library constructs to express parallelism by submitting job passing functions, and general execution information as parameters. The library is meant to overcome limitations imposed by the GIL when using Python's built-in threading library, and exploit computation clusters. The authors claim automatic discovery of computational resources, and their dynamic allocation. However, the library is available neither on PyPi, nor on Github, only on a website \cite{parallelpython2022}.
Celery is a distributed task queue system, which can be used to complete heavy DS and ML computations, but is also meant to have wider applicability in view to support large business applications. Its main components are its broker and backend. The broker is responsible for managing communication between computation threads, and the backend provides the memory storage for queue management. The cost of this flexibility is that deploying Celery is much more complex that alternative approaches mentioned in this section, thus preferred for complex business logic, but probably not for prototypical DS and ML projects. It is actively maintained, and backed by a large community.
The main feature of Playdoh \cite{rossant2013playdoh} is a parallel and distributed map function, as most libraries in this section. It is mainly oriented towards numerical optimization and Monte-Carlo simulation, with some specialized
functions in this area. Also, if the tasks are made of PyCUDA or CUDA code, Playdoh can distribute the work to several GPUs in parallel.
It is worth noting that Parallel Python and Playdoh are not actively maintained, and not supported by Python 3+ interpreters.
Formerly known as IPython.parallel, Ipyparallel \cite{ipyparallel2022} is a Python library for the development of task-based parallel applications. This package leverages the usage of IPython engines in parallel to run tasks. It has four main components: engine, hub, schedulers, and client. The engine is a subclass of the IPython kernel for Jupyter, and is responsible for running user tasks as commanded by a scheduler. The hub manages the cluster by keeping track of schedulers and clients. With this architecture, Ipyparallel allows to abstract potentially heterogeneous distributed computation facilities, accessed by multiple clients working collaboratively. However, its powerful abstractions requires significant boilerplate code, preventing straightforward adaptation of existing DS and ML projects.

Asynchronous function execution is central in Parsl \cite{babuji2019parsl}. Its task-based distributed programming model is based on Parsl \emph{apps}, which may be decorated Python functions (\emph{@python\_app}) or calls to shell commands (\emph{@bash\_app}). Like for torcpy \cite{HADJIDOUKAS2020100517} and SCOOP \cite{Geoffroy2014}, task distribution is enabled by a custom \emph{futures} implementation to manage asynchronous function execution in a distributed context. Its specificity is to facilitate function chaining, which enables parallelization of complex jobs. An Executor is deployed on each host of a distributed architecture, each managing several local workers. Available resources and task distribution are abstracted by a Data Flow Kernel on the client side. A set of launchers accommodate for various high-performance cluster types. Parsl is backed by an active community.

Ray \cite{moritz_ray_2018} is a versatile task-based parallelization tool. On the one hand, it provides low-level facilities, based on primitives like actors and tasks that allow to define and manage distributed computations. Tasks are stateless and executed asynchronously, while actors represent stateful computations. On the other hand, it also provides high-level libraries for deep and reinforcement learning, data processing and analytics. Ray is widely adopted, serving as a parallel framework for other libraries like Modin, LightGBM, and Mars. While it provides direct support for frameworks such as Tensorflow and PyTorch, it can also act as a communication layer for Horovod. Ray is straightforward to install and get up and running in its simplest configuration (multi-core CPU), but requires investment for complex configurations with multiple computation nodes, each possibly holding multiple GPUs. It is therefore rather meant for practitioners with production needs.

Dace \cite{dace} is a Python library that translates Python code to C++ using the \emph{@dace} decorator. It transforms the code into a Stateful DataFlow multiGraph (SDFG), supporting a subset of Python code, Numpy operators, and explicit data flows. The SDFG is a directed graph where nodes represent containers or computations, and edges indicate data movement. Dace has two types of containers: data (memory-mapped arrays) and stream (concurrent queues). Computation containers contain stateless functions. The Python to C++ compiler in Dace leverages the Python AST to infer types, shapes, and perform code analysis. SDFGs enable parallelism by grouping subgraphs, and optimizations are applied through graph transformations. Compilation involves inferring data dependencies, hierarchical code generation, and invoking the compiler for the desired output. While basic usage is simple on multi-core CPUs, complex scenarios require additional development. Dace heavily relies on the host software environment and may be harmed by version clashes. GPU and FPGA support is possible but requires advanced knowledge of the library.

Dask~\cite{rocklin2015dask, dask2016} is a widely used task-based distributed computing library closely integrated with Numpy and Pandas. While it shares a similar API with these libraries, and could arguably be considered as a drop-in library (see Section \ref{sec:libraries}), it introduces task-based logic and incurs setup overheads. Dask provides APIs for arrays (similar to Numpy), data frames (similar to Pandas), and lists (similar to Python iterators). It utilizes task schedulers to split arrays or dataframes into smaller pieces, distribute work, and merge results. Computation is represented as a directed acyclic graph (DAG) where tasks are defined as function-argument tuples and can be executed concurrently. Dask supports various task schedulers for single or multiple nodes in a cluster, allowing customizable configuration. Although Dask does not directly support GPUs, it can schedule GPU-related work at the task level using Dask-cuDF, which extends the cuDF dataframe library (see Section \ref{sec:libraries}) within Dask.

The Tuplex library \cite{Spiegelberg2021} exhibits a similar API to Dask and utilizes optimized LLVM bytecode generation to achieve maximum acceleration. Tuplex performs a dynamic analysis, by considering both the code and the data for code generation. Impressive results have been reported, demonstrating up to 91x acceleration in intricate data pre-processing tasks involving User-Defined Functions encompassing operations like regular expressions and query joins. While the present paper primarily focuses on DS and ML jobs with intense numerical computations, it is noteworthy that complex pre-processing and business tasks can assume critical importance in the management of large-scale data within production environments. Therefore, Tuplex remains a relevant tool, although its improvements may not be considered spectacular in the context of typical research ML projects.

\section{Discussion and Results}
In this article we have presented numerous tools and techniques that are proposing enhancement of performances of Python in the context of DS and ML.
We have tried to depict those tools from the perspective of practitioners, in order to provide them with sufficient insights to select and use an appropriate tool in this still-ongoing quest for Python performance enhancement. We thus have infused the need of practitioners into stereotypical scenarios and assigned existing tools and approaches to the most relevant scenario at hand.

\subsection{Results}

We have identified different kind of techniques during our survey that we shortly summarize here: 
\begin{itemize}
	\item Parallelization libraries: \textbf{MPI}, \textbf{OpenMP} and \textbf{Task-based},
	\item Drop-in libraries,
	\item Program transformation: \textbf{transpilers}, \textbf{JIT}, \textbf{General Compilers} (e.g., LLVM based),
    \item Complete frameworks.
\end{itemize}

\subsubsection{First scenario: pure Python performance improvement} \label{sec:firstsummary}

The tools relevant to this user scenario are summarized in Table~\ref{tab:scenario1summary}. In this table, the surveyed tools are characterized by:

\begin{enumerate}
	\item Tool name and reference,
	\item The implementation technique for performance enhancement (based on the aforementioned list of techniques),
	\item Supported acceleration on CPU, GPU or both,
	\item Usage complexity: this is a qualitative judgement of the effort to put in the effective usage of the tool. It combines an evaluation of the time needed to master its API, its impact on the structure of an existing piece of code, and the efforts needed for its deployment. The number of + denotes the complexity, getting 3 + means that the tool is complex to learn and potentially intrusive in code and may require a lot of tweaks. Getting a - means that the tool requires little work beyond few command lines or editing a configuration file, for example,
	\item Any additional limitation or requirement.
\end{enumerate}

The first scenario assumes the existence of a pure Python codebase, which must be accelerated and parallelized. Therefore, it is mainly relying on parallelization libraries and program transformations. However, due to their genericity, some of the tools described in this scenario could also apply to other scenarios.  Notably, some tools are already applicable for the enhancement of performances of specialized DS libraries (e.g., ALPyNa, Numba).
As we can see in Table~\ref{tab:scenario1summary}, some of the tools rely on task-based parallelization behind the scenes. Using the latter as a structuring framework generally comes with technical complications (see Section \ref{sec:distributed}), but the tools surveyed in this section scaffold this complexity as much as possible. Alternatively, some of the proposed tools act as wrappers from existing C/C++ libraries already offering great performances.

Transpilation and compilation approaches offer to hide some of the complexity for the practitioner. The simpler ones do not require anything from the practitioner, except doing the compilation. The most advanced ones are relying on code annotation to guide the compilation to perform acceleration and parallelization.
In general, all those approaches require more involvement of the practitioner to make them work, being potentially quite intrusive on the code through high refactoring (e.g., MPI based techniques).
Finally, very few propose to exploit a GPU, as it is known as a complex case for general purpose programming.

\begin{landscape}
\begin{table}[h]\centering\scriptsize
\begin{tabular}{p{1.8cm}p{2.3cm}p{2.2cm}p{2.2cm}p{2.2cm}p{2.2cm}p{2.2cm}p{2.2cm}}

\toprule

 \textbf{Tool Name} & 
 \textbf{Activity Period} & 
 \textbf{Technique} & 
 \textbf{Hardware} & 
 \textbf{Usage Complexity} & 
 \textbf{Open Source} & 
 \textbf{Maintained by} & 
 \textbf{Popularity}  \\ \hline

\midrule 
  		
 \textbf{MPI4Py} \cite{DALCIN20051108} & 10.2006 - 11.2022 $\star$ & MPI & CPU & +++ & Y & Individuals  & 632 / 174,734  \\
 \textbf{Comments} & \multicolumn{7}{p{17.6cm}}{Used by other libraries and software such as H5py, vtk, dask, and PcoketFlow. DS/ML practitioners may prefer using a library that hides low-level MPI instructions to obtain parallelism.}  \\ \hline
 
 \textbf{PyPar} \cite{pypar2022} & 06.2007 - 11.2016 $\ast$ & MPI & CPU & +++ & Y & Individuals & 68 / na \\
  \textbf{Comments} &  \multicolumn{7}{p{17.6cm}}{Used by other projects such as ANUGA, TCRM, and Wind multipliers. Helps to avoid the construction of memory buffers by using Pickle. Not compatible with Python 3+.} \\ \hline 

 \textbf{Pymp} \cite{pymp} & 05.2016 - 03.2022 $\star$ & OpenMP  & CPU & ++ & Y & Individuals & 256 / 66,880 \\
  \textbf{Comments} &  \multicolumn{7}{p{17.6cm}}{Only work on systems with fork support. Overhead to circumvent the GIL by using the OS fork method.} \\ \hline 

 \textbf{PyOMP} \cite{PyOMP2021} & 06.2021 - 01.2023 $\ast$ & OpenMP & CPU & ++	& Y & Individuals & 33 / na \\
  \textbf{Comments} &  \multicolumn{7}{p{17.6cm}}{Only works with Linux x86-64 with one specific Python and NumPy version. It is distributed as a forked Numba package with a version out of the main branch.} \\ \hline 
	
 \textbf{PyCOMPSs} \cite{Tejedor2017} & 01.2017 - 11.2022 $\star$ & Task-based & CPU & + 	& Y & University & na / 96 \\
  \textbf{Comments} &  \multicolumn{7}{p{17.6cm}}{Wrapper for COMPS \cite{Tejedor2008}. Used mostly in academic environments. The user needs to know task-based programming and use the definitions of the library.} \\ \hline 	
  
 \textbf{Pygion} \cite{slaughter2019pygion} & na & Task-based & CPU & + & na & na & na \\
  \textbf{Comments} &  \multicolumn{7}{p{17.6cm}}{Wrapper for Legion \cite{bauer2014legion}.}  \\ \hline

 \textbf{PyKokkos} \cite{AlAwarETAL21PyKokkos}	& 04.2021 -	08.2022 $\ast$ & Task-based & CPU & + & Y & Individuals	& 58 / 41 \\
  \textbf{Comments} &  \multicolumn{7}{p{17.6cm}}{Wrapper for Kokkos \cite{Trott2022}. Works only with Ubuntu with gcc 7.5.0 and NVCC 10.2.} \\ \hline
  
 \textbf{Jug} \cite{coelho2017jug} & 07.2019 - 05.2022 $\star$ & Task-based & CPU & ++	& Y & Individuals & 390 / 567 \\
  \textbf{Comments} &  \multicolumn{7}{p{17.6cm}}{The user requires familiarity with task-based programming and library definitions.} \\ \hline
  
 \textbf{Pydron} \cite{Muller2014}	& 01.2016 - 01.2016 $\star$ & Task-based & CPU & ++ & Y & Individuals & 5 / 1 \\
  \textbf{Comments} &  \multicolumn{7}{p{17.6cm}}{The user requires familiarity with task-based programming and library definitions.} \\ \hline
 
  \textbf{Cython}	& 08.2007 - 01.2023 $\star$ & Compiler & CPU & +++	& Y & Individuals & 7.9k / 30,221,789 \\
  \textbf{Comments} &  \multicolumn{7}{p{17.6cm}}{To easily create a C extension for Python, including parallel processing. Used by many popular libraries such scipy, pandas, and numpy. Requires manual refactoring of code for performance improvement.} \\ \hline

  \textbf{Numba} \cite{numba2015}	& 12.2017 - 12.2022 $\star$ & JIT & Both & +	& Y & Enterprise & 8.5k / 2,321,402 \\
  \textbf{Comments} &  \multicolumn{7}{p{17.6cm}}{According to the statistics it is a highly popular package. It uses decorators to give hints to the compiler and has a dependency on LLVM.} \\ \hline

  \textbf{Hope} \cite{AKERET20151}	& 10.2014 - 05.2018 $\star$ & Transpilation & CPU & +/- & 	Y & University & 384 / 167 \\
  \textbf{Comments} &  \multicolumn{7}{p{17.6cm}}{Used mainly in academic environment as support for other packages such as PyCosmo. Only supports a subset of Python.} \\ \hline
  
  \textbf{Shed skin} \cite{shedskin2022} & 07.2009 - 03.2023 $\ast$ & Transpilation & CPU & ++	& Y & Individuals & 673 / 35 \\
  \textbf{Comments} &  \multicolumn{7}{p{17.6cm}}{All variables are implicitly typed. Only supports Python  2.4-2.6.} \\ \hline
  
  \textbf{Nuitka} \cite{nuitka2022}	& 02.2013 - 01.2023 $\star$ & Transpilation & CPU & -	& Y & Individuals & 8.7k / 66,296 \\
  \textbf{Comments} &  \multicolumn{7}{p{17.6cm}}{Translates Python code to highly optimized C or C++ code. Used by at least other 25 packages.} \\ \hline
  
  \textbf{Pythran}	\cite{Guelton2015}	& 08.2012 - 01.2023 $\star$ & Compilation & CPU & -	& Y & Individuals & 1.9k / 262,749 \\
  \textbf{Comments} &  \multicolumn{7}{p{17.6cm}}{Used by at least 13 packages on Github to accelerate computations. Generates C++ code and optimizes it. The code needs to be compiled to be used in another project as a module.} \\ \hline
  
\textbf{Transpyle}	& 12.2017 - 08.2019 $\star$ & Transpilation & CPU & + (annotations) & Y & Individuals & 122 / 78 \\
  \textbf{Comments} &  \multicolumn{7}{p{17.6cm}}{Can bridge Python, Fortran and C/C++. Only a subset of the language is supported.} \\ \hline

  \textbf{Autoparallel} \cite{Ramon-Cortes2020}	& 11.2017 - 08.2020 $\ast$ & Compilation & CPU & +	& Y & Individuals & 11 / na \\
   \textbf{Comments} &  \multicolumn{7}{p{17.6cm}}{Relies on PyCOMPSs and PLUTO. The code is annotated and then automatically translated into PyCOMPSs task definition.} \\ \hline
  
        %ALPyNA \cite{alpyna2019}	\newline na	& na	\newline na \newline na & - na \newline - na \\ \hline
  \textbf{Pyjion} \cite{pyjion2023} & 11.2020 - 11.2022 $\star$ & JIT & CPU & +++ & Y & Individuals & 1.3k / 894 \\
  \textbf{Comments} &  \multicolumn{7}{p{17.6cm}}{Compiles CPython bytecode into improved machine code. Does not currently support with blocks and async...await (YIELD\_FROM) statements.} \\ \hline

 \textbf{Pyston-lite} \cite{pyston2022} & 03.2021 - 03.2023 $\ast$& JIT & CPU & - & Y & Enterprise & 2.4k / 272,262 \\
  \textbf{Comments} &  \multicolumn{7}{p{17.6cm}}{JIT extension to CPython. Usage is straightforward, but not frequently updated on PyPI. Needs to be built for a recent version, soon to be incorporated to CPython.} \\ \hline

\bottomrule   
\end{tabular}
	\begin{flushleft}
Activity period format: month.year. $\star$ First and last release on PyPI $\cdot$ $\ast$ First and last commit on Github	$\cdot$ $\pm$ First and last update on sourceforge	\newline
Popularity: Github stars / PyPI last month downloads. Statistics as as of March 2023.
	\end{flushleft}
	\caption{Summary information of the tools from first scenario}
	\label{tab:scenario1summary}
\end{table}

\end{landscape}
% Summary on how complex it is to write code

\subsubsection{Second scenario: Accelerating numerical libraries usage} 

The tools relevant to this user scenario are summarized in Table~\ref{tab:scenario2summary}. In this context, it is assumed that the existing codebase relies on one of the most commonly used computation libraries: Numpy, Pandas or Scikit-learn. Tools and approaches in this section aim at enhancing or replacing these libraries.

In Table~\ref{tab:scenario2summary}, we can see that most of the found approaches are drop-in libraries that replace as much as possible the syntax of the original library, keeping the same semantic but providing enhancement. 
Their usage is sometimes as simple as function call substitution.
A few tools provide the exploitation of GPU devices for performance acceleration. 
For maximal benefits, they require additional operations relating to memory movement between central and GPU memory.
In the context of CuPy, it materializes as copying Numpy arrays in CuPy ones.
Like Scikit-learn relies on Numpy and Pandas, cuML relies on CuPy and cuDF to offer a broad coverage of the former. 
Many drop-in alternatives exist for Numpy, which is explained by the very high popularity of Numpy as a building block for DS and ML code development, and as a dependency in other Python libraries.

\begin{landscape}
\begin{table}[h]\centering\scriptsize
\begin{tabular}{p{1.9cm}p{2.3cm}p{2.2cm}p{2.2cm}p{2.2cm}p{2.2cm}p{2.2cm}p{2.1cm}}

\toprule

 \textbf{Tool Name} & 
 \textbf{Activity Period} & 
 \textbf{Library / Technique} & 
 \textbf{Hardware} & 
 \textbf{Usage Complexity} & 
 \textbf{Open Source} & 
 \textbf{Maintained by} & 
 \textbf{Popularity}  \\ \hline

\midrule 

 \textbf{Distarray} \cite{distarray2022} & 04.2014 - 10.2015 $\star$ & Numpy / Drop-in & CPU & + & Y & Individuals	& 4 / 15 \\
 \textbf{Comments} & \multicolumn{7}{p{17.6cm}}{Distributed Numpy arrays based on MPI. No updates since 2015.} \\ \hline 

 \textbf{DistNumPy} \cite{Kristensen2010} & 04.2008 - 09.2011 $\ast$ & Numpy / Drop-in	& CPU & -	& Y & Individuals & 5 / na \\
 \textbf{Comments} & \multicolumn{7}{p{17.6cm}}{The project is deprecated, moved to Bohrium.} \\ \hline 

 \textbf{Bohrium} \cite{kristensen2013bohrium} & 11.2017 - 11.2020 $\star$ & Numpy / Drop-in & Both & +	& Y & Individuals & 218 / 336 \\
 \textbf{Comments} & \multicolumn{7}{p{17.6cm}}{Bohrium is used as support for other packages such as Veros and Weld. Drop-in replacement for Numpy but with limited coverage, claimed fall back to Numpy, observed crashes experimentally.} \\ \hline 
		
 \textbf{D2O} \cite{steininger2016d2o} & 05.2016 - 09.2017 $\mp$ & Numpy / Drop-in & CPU & + & Y & Individuals & na / na \\
 \textbf{Comments} & \multicolumn{7}{p{17.6cm}}{Last update since 2017. Manual-refactoring is needed to use the distributed data structure.} \\ \hline 
 
 \textbf{CuPy} \cite{nishino2017cupy} & 06.2016 - 01.2023 $\star$ & Numpy / Drop-in & GPU & + & Y & Enterprise & 6.9k / 22,849 \\
 \textbf{Comments} & \multicolumn{7}{p{17.6cm}}{Good Numpy coverage but not complete. Used by other libraries and part of the RAPIDS ecosystem. The user must add instructions for data copies between the GPU memory and main memory. CuPy has many packages for different CUDA and AMD ROCm versions.} \\ \hline 

 \textbf{PyPacho} \cite{pypacho2021} & 04.2018 - 02.2021	& Numpy / Drop-in	 & Both & ++ & Y & Individuals &  1 / na \\
  \textbf{Comments} & \multicolumn{7}{p{17.6cm}}{Numerical computations based on OpenCL and CUDA. Not documented.}  \\ \hline 
 
 \textbf{DelayRepay} \cite{Morton2020} & xx.2020 - xx.2020	& Numpy / Drop-in	 & Both & - & na & na & na \\
  \textbf{Comments} & \multicolumn{7}{p{17.6cm}}{Method remains conceptual as no associated code can be found.}  \\ \hline 
 
 \textbf{PyViennaCL} \cite{pyviennacl2022} & 03.2014 - 05.2014 $\star$ & Numpy / Drop-in & Both & - & Y & Individuals & 32 / 27 \\
  \textbf{Comments} & \multicolumn{7}{p{17.6cm}}{Provides functions to create data structures (Vector and Matrix) and apply mathematical operations on multi-CPUs and GPUs.} \\ \hline 
 
 \textbf{JAX} \cite{jax2018github} & 12.2018 - 01.2023 $\star$ & Numpy / Drop-in/JIT	 & Both & -/++	& Y & Enterprise & 23,000 / 4,135,872 \\
 \textbf{Comments} & \multicolumn{7}{p{17.6cm}}{Uses Accelerated Linear Algebra compiler (XLA) \cite{xla2020}. Used by many packages such as ColabDesign, trax, ml-workspace, and TensorNetwork. It can be used as drop-in library for Numpy. Nonetheless, JAX arrays are always immutable. Additionally, JAX is aimed to work best with functional programming.} \\ \hline 
 
 \textbf{NumExpr} \cite{numexpr2022} & 01.2009 - 10.2022 $\star$ & Numpy / Library & Both & + & Y & Individuals & 2k / 2,975,123 \\
  \textbf{Comments} & \multicolumn{7}{p{17.6cm}}{Used by other popular packages such as Pandas, zipline, and osmnx. The API is not as extensive as Numpy. Requires to work with large arrays to overcome the overhead of compilation and usage.}  \\ \hline 
  
		%ALPyNA \cite{alpyna2019} & \textbf{REPEATED} & na \\ \hline 
		%Numba \cite{numba2015} & \textbf{REPEATED} & na \\ \hline
		%Pythran	\cite{Guelton2015} & \textbf{REPEATED} & na \\ \hline
		%Jug \cite{coelho2017jug} & \textbf{REPEATED} & na \\ \hline
		
 \textbf{Vaex} \cite{breddels2018vaex} & 02.2015 - 12.2022 $\star$ &  Pandas / Drop-in 	 & CPU & -/+	& Y & Individuals & 7.9k / 52,307 \\
  \textbf{Comments} & \multicolumn{7}{p{17.6cm}}{Used by at least 33 Github packages such as neural-lifetimes, geospatial-ml, radis, and optimus. This tool prioritizes memory optimization for handling large datasets, which may impact performance.} \\ \hline 
 
 \textbf{Modin} \cite{petersohn2020towards} & 06.2018 - 01.2023 $\star$	& Pandas / Drop-in 	 & CPU & -  & Y & Individuals & 8.6k / 1,390,926 \\
 \textbf{Comments} & \multicolumn{7}{p{17.6cm}}{Can use execution engines like Dask \cite{rocklin2015dask} or Ray \cite{moritz2018ray}. Used by at least 47 Github packages such as aws-sdk-pandas, ludwig, and pandera. Modin lacks full implementation of certain Pandas functions, invoking unsupported functions incurs overhead as execution falls back to Pandas and results are transferred back to Modin, requiring data type transformation.} \\ \hline 
 
 \textbf{cuDF} \cite{cudf2022} & 11.2018 - 02.2023 $\mp$ & Pandas / Drop-in 	 & GPU & -/+ & Y & Enterprise & 5.5k /  2,212 \\
 \textbf{Comments} & \multicolumn{7}{p{17.6cm}}{It is part of rapids.ai a collection of software. cuDF is restricted to use only with NVIDIA GPUs. It can lead to certain memory constraints. Specifically, due to the relatively smaller size of GPU memory, users may encounter overheads related to memory swaps or copies, which can impact performance.} \\ \hline 
 
 \textbf{datatable} \cite{datatable2022} & 03.2018 - 07.2021 $\star$ & Pandas / Library 	 & CPU & + & Y & Individuals & 1.7k / 73,586 \\
 \textbf{Comments} & \multicolumn{7}{p{17.6cm}}{Used by at least three other packages in PyPI.} \\ \hline 
 
 \textbf{polars} \cite{polars2022} & 03.2021 - 03.2023 $\star$ & Pandas	/ Library 	 & CPU & +	& Y & Enterprise & 16.9k / 466,445 \\
 \textbf{Comments} & \multicolumn{7}{p{17.6cm}}{Used by 186 packages and 1,664 repositories on Github. There are some differences between the Pandas API and the polars API.} \\ \hline 
		
 \textbf{Dislib} \cite{dislib2019} & 02.2019 - 11.2022 $\star$ & SciKit	/ drop-in 	 & CPU & ++ & Y & University & 41 / 411 \\
 \textbf{Comments} & \multicolumn{7}{p{17.6cm}}{Provides a set of distributed algorithms (related to ML and data processing) such as regression techniques, K nearest neighbors, K-means. The library has been implemented on top of PyCOMPSs programming model.} \\ \hline 
 
 \textbf{cuML} \cite{rapids2022} & 01.2019 - 06.2020 $\star$ & SciKit / Library  & GPU & ++ & Y & Enterprise & 220 / 2,059 \\
 \textbf{Comments} & \multicolumn{7}{p{17.6cm}}{Compatible with CuPy for Numpy support. Used by at least 6 packages and 178 repositories on Github. Does not convert code to run into GPUs, helps with deployment and management of Dask workers.} \\ \hline 
 
 \textbf{MLlib} \cite{meng2016mllib} & 10.2012 - 04.2023 $\mp$ & SciKit	/ Library 	 & CPU & + 	& Y & Enterprise & na / na \\
 \textbf{Comments} & \multicolumn{7}{p{17.6cm}}{Must be used within Spark. Part of Spark, installed as pyspark. MLlib is part of spark, and in PyPI it is named pyspark.} \\ \hline 

\bottomrule   
\end{tabular}
	\begin{flushleft}
Activity period format: month.year. $\star$ First and last release on PyPI $\cdot$ $\ast$ First and last commit on Github	$\cdot$ $\pm$ First and last update on sourceforge	\newline
Popularity: Github stars / PyPI last month downloads. Statistics as as of March 2023.
	\end{flushleft}
	\caption{Summary information of the tools from second scenario}
	\label{tab:scenario2summary}
\end{table}

\end{landscape}

\subsubsection{Third scenario: structuring frameworks} 

The tools relevant to this user scenario are summarized in Table~\ref{tab:scenario3summary}. This section surveyed tools which deeply affect an existing codebase, and thus should preferably be used right when the implementation of a DS or ML algorithm start. As a counterpart, they generally provide many primitives which facilitate the work of the practitioner if they stick to the framework driving principles. We framed deep learning frameworks in this category, as they come with their very own logic to which the data scientist must adapt. In exchange from this effort, they come with high-level abstractions, and scaffold the access to GPU hardware so that maximal performance is obtained with minimal specific development effort.

In this section, we also gathered distributed computing frameworks. They generally have wider applicability compared to deep learning frameworks, and sometimes act as back-end for tools summarized in Section \ref{sec:firstsummary}. However, when used in first intention, they come with specific code constructs which heavily constrain software development, as well as complex setup procedures to deal with variable cluster configurations. As a consequence, it is generally better to involve these tools when implementation starts. Using these frameworks then pays off in terms of the size of the data sets they can handle, which can be orders of magnitude larger than with other tools surveyed elsewhere in this article.

\begin{landscape}
\begin{table}[h]\centering\scriptsize
\begin{tabular}{p{2.3cm}p{2.3cm}p{2.2cm}p{2.1cm}p{2.1cm}p{2.1cm}p{2.2cm}p{2.1cm}}

\toprule

 \textbf{Tool Name} & 
 \textbf{Activity Period} & 
 \textbf{Techniques} & 
 \textbf{Hardware} & 
 \textbf{Usage Complexity} & 
 \textbf{Open Source} & 
 \textbf{Maintained by} & 
 \textbf{Popularity}  \\ \hline

\midrule 

 \textbf{Tensorflow} \cite{tensorflow2015-whitepaper} & 12.2016 - 03.2023 $\star$ & Framework, JIT & Both & +++ & Y & Enterprise & 175k / 15,397,352 \\
 \textbf{Comments} & \multicolumn{7}{p{17.6cm}}{Used by 4,934 packages and 275,967 repositories on Github. Powerful and widely adopted, steep learning curve, industry standard.} \\ \hline
 
 \textbf{Keras} \cite{chollet2015keras} & 11.2015 - 03.2023 $\star$ & Library & Both & + & Y & Enterprise & 58k / 10,078,784 \\
 \textbf{Comments} & \multicolumn{7}{p{17.6cm}}{Formerly supported several frameworks, bound to Tensorflow since 2019.} \\ \hline

 \textbf{PyTorch} \cite{pytorch} & 12.2018 - 03.2023 $\star$ & Framework, JIT & Both & ++ & Y & Enterprise & 67.2k / 9,992,796 \\
 \textbf{Comments} & \multicolumn{7}{p{17.6cm}}{Used by 225,826 repositories on Github. Major deep learning framework, high flexibility, more popular in research and academic contexts.} \\ \hline

 \textbf{Theano} \cite{theano} & 11.2010 - 07.2020 $\star$ & Framework & Both & ++ & Y & University & 9.7k / 229,250 \\
 \textbf{Comments} & \multicolumn{7}{p{17.6cm}}{Used by 203 packages and 13,754 repositories on Github. Former competitor of Tensorflow, development stopped in 2020. Forked by Aesara \cite{aesara} since.} \\ \hline 

 \textbf{MXNet} \cite{mxnet} & 05.2017 - 03.2022 $\star$ & Framework, JIT & Both & +++ & Y & Foundation & 20.4k / 351,957 \\
 \textbf{Comments} & \multicolumn{7}{p{17.6cm}}{Used by 94 packages and 6,381 repositories on Github. Similar to PyTorch in terms of flexibility, lacks convenient IO primitives, and the release frequency has slowed down recently.} \\ \hline

 \textbf{TorcPy} \cite{HADJIDOUKAS2020100517} & 10.2019 - 12.2021 $\ast$ & Task-based & CPU & ++ & Y & Enterprise & 16 / na \\
 \textbf{Comments} & \multicolumn{7}{p{17.6cm}}{Extending Python built-in CPU parallelization to cluster environments, but has little adoption.} \\ \hline
 
 \textbf{Horovod} \cite{sergeev2018horovod} & 08.2017 - 02.2023 $\star$ & Task-based, MPI & GPU & + & Y & Enterprise & 13.3k / 71,025 \\
 \textbf{Comments} & \multicolumn{7}{p{17.6cm}}{Used by 22 packages and 892 repositories on Github. Distributed deep learning framework based on OpenMPI or Gloo for communication.} \\ \hline 
 
 \textbf{Charm4py} \cite{Galvez2018} & 11.2018 - 09.2019 $\star$ & Task-based & CPU & ++ & Y & University & 280 / 162 \\
 \textbf{Comments} & \multicolumn{7}{p{17.6cm}}{Task-based distributed computing framework with distributed Python objects. Note that Python 3+ is not supported.} \\ \hline 
 
 \textbf{SCOOP} \cite{Geoffroy2014} & 01.2013 - 08.2022 $\star$ & Task-based & CPU & ++ & Y & University & 591 / 1,415 \\
 \textbf{Comments} & \multicolumn{7}{p{17.6cm}}{Used by 6 packages and 326 repositories on Github. Similar to a MapReduce framework, with extensive information available for deployment on high-performance clusters.} \\ \hline 
 
 \textbf{Parallel Python} \cite{parallelpython2022} & na - na $\mp$ & Task-based & CPU & ++ & Y & na & na / na \\
 \textbf{Comments} & \multicolumn{7}{p{17.6cm}}{Job-based distributed computing framework. It is absent from main repositories, and Python 3+ is not supported.} \\ \hline 

 \textbf{Celery} \cite{Celery2022} & 04.2009 - 02.2023 $\star$ & Task-based & CPU & +++ & Y & Enterprise & 21.5k / 6,198,447 \\
 \textbf{Comments} & \multicolumn{7}{p{17.6cm}}{Used by 1,515 packages and 102,009 repositories on Github. Popular distributed task queue system with many features, but can be complex to deploy.} \\ \hline

 \textbf{Playdoh} \cite{rossant2013playdoh} & 03.2010 - 02.2011 $\star$ & Task-based & Both & ++ & Y & University & 67 / 17 \\
 \textbf{Comments} & \multicolumn{7}{p{17.6cm}}{Distributed map function that also supports CUDA code. Not maintained since 2011, and does not support Python 3+} \\ \hline

 \textbf{Ipyparallel} \cite{ipyparallel2022} & 04.2015 - 03.2023 $\star$ & Task-based & CPU & ++ & Y & University & 2.4k / 143,330 \\
 \textbf{Comments} & \multicolumn{7}{p{17.6cm}}{Powerful but requires significant code adaptation.} \\ \hline

 \textbf{Parsl} \cite{babuji2019parsl} & 02.2017 - 03.2023 $\star$ & Task-based & CPU & ++ & Y & University & 376 / 416,826 \\
 \textbf{Comments} & \multicolumn{7}{p{17.6cm}}{Task-based parallelization library that uses asynchronous function invocation, backed by an active community.} \\ \hline

 \textbf{Tuplex} \cite{Spiegelberg2021} & 06.2021 - 10.2022 $\star$ & Task-based, Compil. & CPU & ++ & Y & University & 798 / 90 \\
 \textbf{Comments} & \multicolumn{7}{p{17.6cm}}{Task-based and compilation-based framework for CPU parallelism. Requires the use of its own Domain Specific Language (DSL).} \\ \hline

 \textbf{Dask} \cite{rocklin2015dask, dask2016} & 01.2015 - 03.2023 $\star$ & Task-based & Both & +/++ & Y & Enterprise & 10.9k / 7,016,339 \\
 \textbf{Comments} & \multicolumn{7}{p{17.6cm}}{Used by 1,583 packages and 47,013 repositories on Github. Supports both single-machine and distributed computing. Strongly bound to Numpy and Pandas.} \\ \hline

 \textbf{Ray} \cite{moritz_ray_2018} & 06.2017 - 03.2023 $\star$ & Task-based, Backend & Both & +++ & Y & Enterprise & 25.8k / 2,095,907 \\
 \textbf{Comments} & \multicolumn{7}{p{17.6cm}}{Used by 553 packages and 9,355 repositories on Github. Task-based framework with a versatile backend supporting both single-machine and distributed computing. Powerful but can be complex to set up effectively.} \\ \hline

 \textbf{Dace} \cite{dace} & 03.2019 - 03.2021 $\star$ & Task-based & Both & +++ & Y & University & 378 / 8,348 \\
 \textbf{Comments} & \multicolumn{7}{p{17.6cm}}{Used by 3 packages and 6 repositories on Github. Supports both single-machine and distributed computing. Powerful capabilities, but requires significant work overhead for complex use cases and in a GPU context.} \\ \hline

\bottomrule   
\end{tabular}
	\begin{flushleft}
Activity period format: month.year. $\star$ First and last release on PyPI $\cdot$ $\ast$ First and last commit on Github	$\cdot$ $\pm$ First and last update on sourceforge	\newline
Popularity: Github stars / PyPI last month downloads. Statistics as as of March 2023.
	\end{flushleft}
	\caption{Summary information of the tools from third scenario}
	\label{tab:scenario3summary}
\end{table}

\end{landscape}

\subsection{Scope and Limitations of the Study}

To mitigate the risk of being biased by our own research we tried to be as open as possible following a simple narrative process. In addition, the narrative review allows us to provide DS and ML practitioners with an overall view on the different existing techniques. It is also sufficiently open to interest practitioners from related areas which make occasional usage of ML techniques, such as scientific computing. 

Performance enhancement of programs is a wide subject including parallelization, and port between architectures and languages. Many tools and approaches exist outside the Python world, and beyond ML and DS. However, to deliver a consistent and organized view on the subject we restrain our subject to cover the three main scenarios that could occur from a data scientist's point of view. Indeed, this is a partial and oriented view on subject leaving space for further explorations.

It is true that software development is a fast-paced domain, and that best options today may be superseded by their competitors or new players only a couple of years after this paper has been issued. However, we believe that the scenarios that back the structure of our study are general enough to remain true even as new tools are developed and existing tools evolve. Also, our study served to highlight few basic facets of tools, to which new or evolving tools can be fairly straightforwardly attached (e.g. related to task-based frameworks in Section \ref{sec:distributed}, or drop-in numerical libraries in Section \ref{sec:libraries}). Therefore, even as time goes, we believe the insights we delineated will provide useful guidance to practitioners for years to come.

% introduce interpreters from a less negative perspective
As previously stated, when we delimited our search scope, we deliberately excluded Python interpreters from our study as they are likely to interact with libraries mostly used in DS and ML domains. Yet there are many contributions in this area, which deeply affect vanilla Python efficiency: we briefly review them below.

\subsection{Python interpreters} \label{sec:interpreters}

% TO COMBINE WITH TEXT BELOW
% argument that facilitates comparisons, leaving aside potential compatibility issues.
% and asserting that interpreter updates improvement could complement those reported otherwise
%Despite bringing visible performance improvements, many Python interpreters, due to the frequent updates to the Python standard, offer only a limited coverage of Python, and more importantly of its libraries.

CPython serves as the primary testbed for new features and language enhancements in Python. This is because it is maintained by the same community that designs the language. The performance challenges encountered in Python are directly associated with the CPython implementation. The article by Zhang et al. \cite{zhang2022quantifying} shows potential performance improvements achievable through different optimization approaches in standard CPython interpreters. These optimizations include techniques like dispatch, branch prediction, and array-style access. 

There are different implementations of the Python interpreter, offering developers the flexibility to replace the default CPython interpreter with alternative options. The advantage of these alternative interpreters is that they can enhance the speed and performance of Python programs without any code modification \emph{a priori}. However, it is important to note that alternative interpreters may have certain limitations and drawbacks that need to be considered.

Alternative Python interpreters include PyPy \cite{PyPySpeed2022}, Pyston \cite{pyston2022},  Cinder \cite{cinder}, and IronPython \cite{ironpython}. However, not all of them offer complete coverage of the Python language. Additionally, some implementations are tied to specific Python versions. Cinder and Pyston, for example, support Python 3.8, IronPython supports Python 3.4, and PyPy supports Python 3.9 (which is relatively closer to the latest CPython versions). Historically, alternative implementations of Python have often lagged behind in terms of supporting the latest language versions and features. 
%This delay in compatibility with the newest Python language versions and futures is a common characteristic of these implementations.

Beyond being bound to Python 3.8, Pyston has no known compatibilities issues\footnote{https://blog.pyston.org/2022/ was accessed on 24/05/2022}. Pyston provides a performance gain estimated between 10 to 35\% on reported benchmarks, despite some overheads observed on specific tasks such as JSON loading\footnote{https://www.phoronix.com/news/Pyston-2.1-vs-Python-3.8-3.9 accessed on 24/05/2022}.
For some time, Pyston has been backed by Anaconda, a popular Python distribution tool. According to recent notes from one of the creators of Pyston, it is likely that their speedups will become fully integrated in a future CPython release\footnote{\url{https://blog.pyston.org/author/kmodzelewski}, consulted on 03/2023}.

Cinder is a JIT interpreter implemented in C++ that includes several performance optimizations such as bytecode inline caching, eager evaluation of coroutines, and a method-at-a-time JIT. According to a report from their maintainers, it can provide a speed up of 1.5 to 4 on many Python performance benchmarks \footnote{https://github.com/facebookincubator/cinder}.
PyPy \cite{PyPySpeed2022} is another alternative Python interpreter featuring a JIT compiler. According to benchmark results, with Pypy the acceleration ratio can range from 0.21 to 4.8\footnote{https://speed.pypy.org/} with respect to CPython, with documented mixed results\footnote{\url{https://pybenchmarks.org/u64q/benchmark.php?test=all\&lang=python3\&lang2=pypy3\&data=u64q}}.
As already mentioned in Section \ref{sec:programtransformation}, the widely used CPython interpreter does not include a built-in JIT compiler. Nonetheless, third-party libraries or tools, such as Pyston-lite (see Section \ref{sec:automatic}) can be utilized to introduce JIT compilation into the Python workflow for performance optimization.

IronPython is an implementation of the Python programming language that runs on the .NET framework, providing seamless integration with .NET technologies and allowing the use of Python libraries. Its performance is comparable to CPython with variations depending on the specific task\footnote{https://www.python.org/about/success/resolver/}. It is important to note that IronPython is primarily designed for Windows environments and does not support the importation of C extension modules, which limits its compatibility with certain libraries like NumPy and SciPy. As a result, some third-party libraries commonly used in DS and ML may not be available in IronPython.

% TO MERGE WITH TEXT ABOVE (MERGED)
A common problem with Python interpreters is that standard libraries - that may depend on other libraries - are not necessarily compliant outside the CPython implementation, and even so, often require building shared libraries from source. This may make it hard to validate the approach for each library and framework, and can be cumbersome for the average practitioner. The Python ecosystem is primarily built around CPython, which means that some community-supported projects, tools, and resources may not work seamlessly, with the added difficulty to keep up with the latest versions and language features.
%which are mainly driven by CPython releases. 

% mention of Julia and other initiatives removed like loopy?

\section{Conclusion}
Our article highlighted different approaches to enhance Python performances regarding three scenarios meant to cover most needs happening in the practice of DS and ML. Each scenario covers a peculiar stereotype of developer dealing with ML and DS tasks. They depict practitioner profiles that range from a very straightforward way of using Python (i.e., vanilla Python), by usage of standard numerical libraries, up to the use of large integrated frameworks. 

By answering our research question, \emph{which approaches can be used to improve Python execution performance in the context of one of these three scenarios?}, we have looked at the most relevant state of the art approaches, following a narrative review principle.
Each scenario calls for specific solutions which may be addressed by different kinds of techniques. For each scenario, we highlighted how given tools may help them deal with their task. We also highlight the estimated complexity to set up those approaches, notably by the impact on the original code and in terms of learning curve. 

We have shown that for pure Python code acceleration, the practitioners have a large choice depending on their level of confidence and control they want to have on the performance improvement. For simple and fast results, but not optimal, they may look at a diverse range of straightforward techniques, some even fully automatic, involving compiler directives, code decorators, or transpilers. Better performance can be obtained with semi-automatic approaches, but they require more involvement from the developer, and a steeper learning curve for maximal gains.

In the case the codebase heavily relies on well-known numerical libraries, the most natural path is to investigate using drop-in libraries. Most of them mimic the API of the library they substitute to, so the learning curve is mild. However, for maximal gains, the practitioner must address subtleties such as memory movements between central and GPU memories.

In the third scenario, the practitioner is starting the development from scratch. Therefore, approaches surveyed in this section are meant to be used right from the start of project development and put heavy constraints of code structure. This initial effort is traded with maximal gains in terms of performance, and minimal surplus of effort if the driving principles of the frameworks are enforced.

We expect this work to give a good comprehensive view and guide practitioners for choosing among the plethora of existing tools. While playing an important role in the structure of our document, we think that this discussion section can also act as a good starting point in this perspective, with all references and then available for deeper inspection. 
Also, all surveyed tools are reported in summary tables, one per user profile, emphasizing distinctive characteristics, and providing usage metrics as to the date this document was written.
Though we tried to be as comprehensive as possible, some features of the surveyed tools may not have been covered. Also, we did not run and quantitatively compare the performance of all the surveyed tools, due to their number and diversity.
It would be almost impossible to find a suitable common benchmark for any Python acceleration method and task dedicated tool. We also expect that our work could help new tool designers who aim at enhancing Python performance to get an overview of the current state of the art.

\bibliographystyle{ACM-Reference-Format}
\bibliography{sample-base}

\end{document}